\documentclass[twocolumn]{autart}    % Enable this line and disable the
\usepackage{graphicx}          % Include this line if your
\usepackage{booktabs}
\usepackage{graphicx,amssymb}
\usepackage{multirow, multicol}
\usepackage{color}
\usepackage{gensymb}
\usepackage{pdflscape}
\usepackage{courier}
\usepackage{epsfig,epstopdf}
\usepackage{booktabs}
\usepackage{chngpage}
\usepackage{mathtools}
\usepackage{longtable}
\usepackage{algorithm}
\usepackage[titletoc]{appendix}
\usepackage{float}
\usepackage{textcomp}
\usepackage{mathrsfs}
\usepackage{array}
\usepackage{subfigure}
\usepackage{amsmath}
\usepackage{caption}
\usepackage{algpseudocode}
\usepackage{color}
\usepackage[round]{natbib}
\usepackage{apalike}
\newtheorem{theorem}{Theorem}
\newtheorem{lemma}{Lemma}

\newtheorem{assumption}{Assumption}

\newtheorem{remark}{Remark}

\begin{document}

\begin{frontmatter}

\title{Bayesian adversarial multi-node bandit for optimal smart grid protection against cyber attacks\thanksref{footnoteinfo}} % Title, preferably not more

\thanks[footnoteinfo]{This work is partially supported by Fudan Scholar and Special Research Grant, UNSW, and by the Australian Research
Council's Discovery Projects funding scheme under Project DP190101566. Corresponding author Huadong Mo. Tel. +61-2-62688683.
Fax +61-2-62688683}

\author[China]{Jianyu Xu}\ead{Jianyu.Xu@xjtlu.edu.cn},    % Add the
\author[UK]{Bin Liu}\ead{b.liu@strath.ac.uk},
\author[Australia]{Huadong Mo}\ead{huadong.mo@adfa.edu.au},               % e-mail address
\author[Australia]{Daoyi Dong}\ead{d.dong@adfa.edu.au}  % (ead) as shown

\address[China]{International Business School Suzhou, Xi'an Jiaotong-Liverpool University, Suzhou, China}  % Please supply
\address[UK]{Department of Management Science, University of Strathclyde, G1 1XQ, Glasgow, UK}
\address[Australia]{School of Engineering and Information Technology, University of New South Wales, Canberra, ACT 2600, Australia}             % full addresses
       % here.

                             % keyword list or with the
                                          % help of the Automatica
                                          % keyword wizard

\begin{abstract}                          % Abstract of not more than 200 words.
The cybersecurity of smart grids has become one of key problems in developing reliable modern power and energy systems. This paper introduces a non-stationary adversarial cost with a variation constraint for smart grids and enables us to investigate the problem of optimal smart grid protection against cyber attacks in a relatively practical scenario. In particular, a Bayesian multi-node bandit (MNB) model with adversarial costs is constructed and a new regret function is defined for this model. An algorithm called Thompson-Hedge
algorithm is presented to solve the problem and the superior performance of the proposed algorithm is proven in terms of
the convergence rate of the regret function. The applicability of the algorithm to real smart grid scenarios is verified and the performance of the algorithm is also demonstrated by numerical examples.
\end{abstract}

\end{frontmatter}

\section{Introduction}
The upgrade of traditional grids to smart grids has brought many benefits to the overall management of power and energy systems, including higher reliability, better efficiency, improved integration of renewable energy resources, more flexible choice for stakeholders and lower operation cost \citep{konstantelos2016strategic,pogaku2007modeling,yu2015analysis}. However, the core technologies, $e.g.$, communication techniques and SCADA systems \citep{abiri2019cyber,khalili2020dis,rana2018smart,todescato2020partition}, which deliver advantages of smart grids, also open the grids to vulnerabilities that already exist in the Information and Communications Technology world. Now, those vulnerabilities pose threats to smart grids, such as denial of service (DoS) attacks, false data injection, replay attacks, privacy data theft and sabotage of critical infrastructure \citep{gallo2020dis,mo2017dynamic,zhu2013distributed}. In addition, the failures in a smart grid caused by cyber attacks can easily cascade to other highly dependent critical infrastructure sectors, such as transportation systems, wastewater systems, health care systems and banking systems, resulting in extensive physical damage and social and economic disruption \citep{abiri2019cyber,che2018cyber}.

While governments, the private sector and academia are recognising the cyber vulnerability of smart grids, the likelihood and impact of a cyber attack are difficult to quantify. Furthermore, for a smart grid, there may be mandatory standards and operation requirements from the grid stakeholders. Current risk management strategies are generally qualitative or heuristic \citep{patsakis2018optimal}. In these strategies, some assumptions, $e.g.$, constant reward with respect to successful anti cyber attack \citep{rana2017cyber,smith2018cyber}, may be unrealistic for most smart grids.

This paper presents a probabilistic risk analysis framework to enhance the smart grid cyber security. In particular, the dynamical and stochastic characteristics of smart grids, such as uncertain demands, are taken into account to investigate the effect of defending strategies on the real operation cost. The Optimal Power Flow model \citep{zhang2015real} is applied to a 11-node radial smart grid originated from the Elia grid, Belgium.
Compared with the existing studies that focus on the inherent risk \citep{mo2017dynamic,zhang2015real}, such as the natural degradation and uncertain renewable energy resources for better maintenance actions and power dispatch, this paper concerns about impact of the external threat - cyber attacks on the operation cost for effective deployment of cyber defense teams. In previous work, the cost of each attack on a node is usually assumed to be a constant \citep{smith2018cyber}. Nevertheless, through investigating some practical scenarios, it has been found that the costs are more likely to be decided by some adversarial factors from the nature. Therefore, an adversarial cost sequence associated with each node is assumed, and a widely used variation constraint is introduced on each cost sequence.
To cope with the objective of sequential decision strategy, the problem is formulated using the reinforcement learning framework \citep{li2020quantum,littman2015reinforcement,sutton2018reinforcement}. Specifically, the Bayesian prior method \citep{smith2018cyber} is employed for the model parameters and the problem is formulated as a Bayesian adversarial multi-node bandit (MNB) model. In addition, a Bayesian minmax type regret function is constructed, which is subject to the learning context.

Research on online algorithms for adversarial MNB problems started from \citet{auer2002nonstochastic}, which is later extended to general adversarial reinforcement learning models by \citep{even2009online}. Recently, \citet{besbes2019optimal} implemented the classical method in \citet{auer2002nonstochastic} for the adversarial bandit problem assuming cost sequences with restricted variation and achieved the state-of-the-art performance for such problems. An obvious gap when using these methods in the problem of our work is that they do not adapt to the Bayesian framework and cannot take advantages of the additional information provided by the Bayesian assumptions. As a result, these algorithms may be ineffective in Bayesian problems in terms of the convergence rate of the regret. An alternative feasible method is proposed in \citet{smith2018cyber}, where model parameters are considered to be system states included in the state space. Therefore, the problem is no longer Bayesian and can be solved using existing algorithms for stochastic MNB. However, this method suffers from a tremendous state space and is always computationally hard even in problems with moderate sizes.

To cope with the abovementioned technical challenge, an online learning method is developed in this study which integrates the Thompson sampling method \citep{russo2014learning} with the classical Hedge algorithm \citep{auer2002nonstochastic,russo2014learning}. Our algorithm takes advantages of both algorithms. In particular, Thompson sampling method is used to cope with the Bayesian framework and Hedge algorithm is applied to the adversarial costs. A theoretical bound on the regret function is proved in the proposed algorithm, which is superior to the typical regret bound by the state-of-the-art algorithms for the same problem. The applicability and numerical performance of the proposed algorithm is also illustrated through real data and simulation studies.

The main contributions of our work are summarized as follows.
\begin{itemize}
\item[1.] A relatively practical adversarial reward with restricted variation constraint is proposed for the MNB model. The prior information of the model parameters is incorporated through a Bayesian framework and a Bayesian adversarial MNB model is formulated.
\item[2.] A Bayesian sup regret is defined as the criterion function for decision making objective of the proposed problem. A new online algorithm, called Thompson-Hedge algorithm, is developed to solve the problem. The convergence rate of the Bayesian sup regret for the proposed algorithm is theoretically proven.
\item[3.] Real smart grid data are used to demonstrate the applicability of the proposed method and numerical results show that the proposed algorithm can achieve the state-of-the-art performance.
\end{itemize}

This paper is organized as follows. Section \ref{section problem formulation} formulates the problem as a Bayesian reinforcement learning model and constructs the regret function. In Section \ref{section thompson hedge algorithm}, an algorithm called Thompson-Hedge algorithm is developed for our learning model and a theoretical upper bound on the regret function is established. Section \ref{section application in smart grid} verifies the feasibility of our model and algorithm using a real data set. In Section \ref{section simulation study}, a comparative study is investigated on the performance between our algorithm and a state-of-the-art algorithm. Section \ref{section concluding remarks} presents concluding remarks.

%%%%%%%%%%%%%%%%%%%%%%%%%%%%%%%%%%%%%%%%%%%%%%%%%%%%%%%%%%%%%%%%%%%%%%%%%%%%%%%%%%%%%%%%%%%%%%
\section{Problem formulation}\label{section problem formulation}
In the electricity network with cyber attacks, an attacker launches a coordinated attack using multiple attack vectors, because the smart grid communication networks are physically distributed and highly heterogeneous. The successful rate of such attack behavior is data-driven, which follows a Poisson distribution, as demonstrated by the empirical study from U.S. Department of Energy \citep{smith2018cyber}. The network defender aims to optimally allocate cyber defense teams among nodes in the network, via probing one node per day. Such a defending action thwarts all attempted cyber attacks to that node on that day, and also helps update his/her belief about the uncertain successful rates of attack. The above interaction between the attacker and the defender has the sequential decision-making nature and leads itself to a Bayesian MNB model. This model employs proactively defense teams that traditionally respond to cyber threats after they occur.

The considered attack scenario in a realistic smart grid is that the DoS attacks block the demand response (DR) messages and dispatch commands \citep{pillitteri2014guidelines,smith2018cyber}. The DoS attacks can be accomplished by flooding the communication channel, $e.g.$, the one between the demand response automation server and customer systems or the one between the control center and power plant router, with other messages and commands, or by tampering with the communication channel \citep{amin2013security,pillitteri2014guidelines,smith2018cyber}. Above actions can prevent legitimate DR messages and dispatch commands being received and transmitted, $i.e.$, depriving authorized access or control to customer systems and power plant, resulting in demand not being responded and power plant not being controlled \citep{pillitteri2014guidelines}. Therefore, the impact of successful DoS attacks on the optimal power flow model of the smart grid is described by making the target node temporarily unavailable and disconnected from the grid, which is illustrated in Section \ref{subsection operation cost of the smart grid}.

Consider a smart grid with $N$ nodes that suffers cyber attacks. Let $\mathbb{N}\triangleq \left\{1,\cdots,N\right\}$ be the set of all nodes. At each time $t=1,2,\ldots$, the operator's action is to choose a node to probe. If $i\in\mathbb{N}$ is probed, the operator observes a (random) number of $K_{i,\,t}$ cyber attacks on node $i$.

The historical cyber incident data from US Department of Energy (Fig. 2 of \citet{smith2018cyber}) has demonstrated that the arrival interval of successful cyber attacks on per node can be well described by a truncated Poisson distribution, and there is no record of multiple cyber attacks (maximum is 3) on per node per day. Therefore, the physical meaning is that in real applications, there should not be more than 3 cyber attacks arriving per node in a defined time interval. In addition, the Palm-Khintchine theorem \citep{heyman2004superposition,smith2018cyber} justifies that the aggregate arrivals from many sources (no need to be strict Poisson) approach a truncated Poisson distribution in the limit time interval. It is assumed that for all $i\in\mathbb{N}$, $\left\{K_{i,\,t},\,t=1,2,\ldots\right\}$ is a sequence of independent and identically distributed random variable sequence drawn from a Poisson distribution truncated at $m>0$ with rate $\lambda_{i}$, $i.e.$, for all $k=1,2,\ldots$, and  $t=1,2,\ldots$,
%
%\begin{align*}
%Pr\left(k_{i,\,t}=k\,|\,\lambda_{i}\right)=\frac{\lambda_{i}^{k}}{k!}e^{-\lambda_{i}}
%,\: \forall t=1,2,\ldots.
%\end{align*}
\begin{align*}
Pr\left(K_{i,\,t}=k\,|\,\lambda_{i}\right)=\left\{
\begin{array}{cc}
\frac{\lambda_{i}^{k}}{k!}e^{-\lambda_{i}}, & k<m \\
\sum_{j=m}^{\infty}\frac{\lambda_{i}^{j}}{j!}e^{-\lambda_{i}}, & k=m\\
0, & \textup{otherwise}
\end{array}
\right..
\end{align*}
In particular, the average number of attacks on each node $i\in\mathbb{N}$, denoted by $\mu_{i}$, is
\begin{equation*}
\mu_{i}=
\sum_{k=1}^{m-1}k\cdot \frac{\lambda_{i}^{k}}{k!}e^{-\lambda_{i}}+
m\sum_{k=m}^{\infty}\frac{\lambda_{i}^{k}}{k!}e^{-\lambda_{i}}.
\end{equation*}
Let $\lambda=(\lambda_{1},\cdots,\lambda_{N})$ and assume a foresight belief on $\lambda$ in terms of a prior distribution $\mathcal{F}=\left(\mathcal{F}_{1},\cdots,\mathcal{F}_{N}\right)$ over $\lambda$, where $\lambda_{i}\sim\mathcal{F}_{i}$ for all $i\in\mathbb{N}$. In particular, it is assumed that each $\mathcal{F}_{i}$ is a gamma distribution with parameters $\left(\alpha_{i},\,\beta_{i}\right)$ and all $\mathcal{F}_{i}$'s are mutually independent with each other. Whenever $k$ cyber attacks are observed at node $i\in\mathbb{N}$, the parameters $\left(\alpha_{i},\,\beta_{i}\right)$ are updated through Bayesian posterior as
$\left(\alpha_{i},\,\beta_{i}\right)\rightarrow \left(\alpha_{i}+k,\,\beta_{i}+1\right)$.

At time $t=1,2,\ldots$, a cost $c_{i,\,t}$ is calculated from an optimization process for all $i\in\mathbb{N}$. Without loss of generality, it is assumed that all $c_{i,\,t}$'s are normalized such that $c_{i,\,t}\in[0,\,1/m]$ for all $i\in\mathbb{N}$ and $t=1,2,\ldots$. The process uses some inputs ($e.g.$, external factors) that are decided by an adversary (environment). Without ambiguity, the cost vector $c_{t}=\left(c_{1,\,t},\cdots,c_{N,\,t}\right)$ is assumed to be decided by the adversary at each time $t=1,2,\ldots$. If node $i\in\mathbb{N}$ is probed, the operator avoids incurring a total cost of
$X(i,\,t)= K_{i,\,t}\cdot c_{i,\,t}$, $X(i,\,t)\in[0,\,1]$. Equivalently, a reward of $K_{i,\,t}\cdot c_{i,\,t}$ is achieved by probing node $\pi_{t}$ at time $t$. An admissible policy $\pi$ is a sequence of mappings $\pi=\left\{\pi_{1},\pi_{2},\cdots\right\}$, where each $\pi_{t}$ is a mapping from historical observations to the set of all probability distributions on $\mathbb{N}$. To distinguish, let $\{i_{1},i_{2},\cdots\}$ represent the sequence of chosen nodes. The axioms of the problem are formulated in the problem protocol below.
%%%%%%%%%%%%%%%%%%%%%%%%%%%%%%%%%%%%%%%%%%%%%%%%%%%%%%%%%%%%%%%%%%%%%%%%%%%%%%%%%%%%%%%%%%%%%%%
\begin{algorithm}[h]
\captionsetup{labelformat = empty}
\renewcommand{\algorithmicrequire}{For each time $t=1,2,\ldots$}
	\caption{\textbf{Problem protocol}: }
	\label{algorithm problem formulation}
	\begin{algorithmic}[1]
		\Require
		\State The adversary decides an cost $c_{i,\,t}$ for all $i\in\mathbb{N}$.
		\State The operator probes a node $i_{t}\in\mathbb{N}$ according to $\pi_{t}$ and observes the number of          cyber attacks on this node to be $k_{i_{t},\,t}$.
		\State All $c_{i,\,t}$'s are revealed to the operator.
		\State A reward $k_{i_{t},\,t}\cdot c_{i_{t},\,t}$ is achieved by the operator.
	\end{algorithmic}
\end{algorithm}

For notational convenience, let $\theta=\left(\alpha_{1},\,\beta_{1},\cdots,\alpha_{N},\,\beta_{N}\right)$ denote the initial parameter vector and $\vec{c}_{T}=\left\{c_{1},\cdots,c_{T}\right\}$ the cost sequence up to time $T$. To proceed, the optimal policy and the regret function is firstly formulated given the parameter vector $\theta$ and cost sequence $\vec{c}_{t}$. The mean reward of node $i\in\mathbb{N}$ at time $t$ is $\mathbf{E}\left(K_{i,\,t}\cdot c_{i,\,t}\right)=\mu_{i}\cdot c_{i,\,t}$. Since the reward sequence from each node is not stationary due to $c_{i,\,t}$, for all $t=1,2,\ldots$, the optimal node is defined as $i_{t}^{*}\triangleq {\arg\max}_{i\in\mathbb{N}}\mu_{i}\cdot c_{i,\,t}$ and a non-stationary optimal policy is supposed to choose node $i_{t}^{*}$ at time $t=1,2,\ldots$. Thus, for any admissible policy $\pi$, given $\theta$ and $\vec{c}_{T}$, the regret function up to time $T$ is defined as
\begin{equation}\label{equation regret}
R^{\pi}\left(\lambda,\,\vec{c}_{T},\,T\right)\triangleq
\sum_{t=1}^{T}\mu_{i_{t}^{*}}\cdot c_{i_{t}^{*},\,t}-
\mathbf{E}^{\pi}\left(\sum_{t=1}^{T}\mu_{\pi_{t}}\cdot c_{\pi_{t},\,t} \,\Big|\, \lambda,\,\vec{c}_{T}\right),
\end{equation}
where $\mathbf{E}^{\pi}$ means the expectation taken with respect to the (random) sequence $\left\{i_{1},\cdots,i_{T}\right\}$ generated by $\pi$.

A constraint is imposed upon the adversary by introducing a sequence of sets
$\left\{\mathbb{C}_{t}\subset [0,\,1]^{t}\right\}_{t\geq2}$, that for all time $t=1,2,\ldots$,
$\vec{c}_{T}\in\mathbb{C}_{t}$. Then, given $\theta$, the regret function with respect to the worst case up to time $T>0$ is
\begin{equation}\label{equation sup regret}
\sup_{\vec{c}_{T}\in\mathbb{C}_{T}} R^{\pi}\left(\lambda,\,\vec{c}_{T},\,T\right).
\end{equation}
Note that $\sup_{\vec{c}_{T}\in\mathbb{C}_{T}} R\left(\lambda,\,\vec{c}_{T},\,T\right)$ is supposed to measure the performance with the action sequence uniformly on all possible cost sequences. Finally, incorporating the prior distribution $\mathcal{F}$ on $\theta$, the Bayesian regret function with respect to the worst case is defined as
\begin{align}\label{equation bayesian regret}
\mathcal{R}^{\pi}\left(\mathbb{C}_{T},\,T\right)\triangleq
\mathbf{E}_{\lambda\sim\mathcal{F}}
\left[\sup_{\vec{c}_{T}\in\mathbb{C}_{T}} R^{\pi}\left(\lambda,\,\vec{c}_{T},\,T\right)\right].
\end{align}
To distinguish between the regret functions in (\ref{equation regret}), (\ref{equation sup regret}) and (\ref{equation bayesian regret}), they are named as regret, sup regret and Bayesian sup regret, respectively, for brevity.

%%%%%%%%%%%%%%%%%%%%%%%%%%%%%%%%%%%%%%%%%%%%%%%%%%%%%%%%%%%%%%%%%%%%%%%%%%%%%%%%%%%%%%%%%%%%%%%%%%%%
By analyzing some real databases (such as the Elia Grid in Section \ref{section application in smart grid}), the sequence of sets $\left\{\mathbb{C}_{T}\subset [0,\,1]^{T}\right\}_{T\geq2}$ is formulated with constraint on the cost. It is concluded from the statistical analysis of the real data that the temporal variations of the cost sequence of each node $i\in\mathbb{N}$, $i.e.$, $\left\{|c_{i,\,t+1}-c_{i,\,t}|\right\}_{t\geq1}$, are stationary and uniformly upper bounded by a value, which only relies on $i$ and is small compared to the average value of $\left\{c_{i,\,t}\right\}_{t\geq1}$. Therefore, it is assumed that for each node the cumulative temporal variation up to time $t\geq1$ is upper bounded by a linear function of $t$. Based on this assumption, a uniform upper bound on the cumulative temporal variation for all nodes is introduced and $\left\{\mathbb{C}_{T}\subset [0,\,1]^{T}\right\}_{T\geq2}$ is constructed as follows.
\begin{align}\label{equation CT}
\begin{split}
\mathbb{C}_{T}\triangleq &\{\left\{c_{i,\,t}\right\}_{t=1}^{T}\in[0,\,1]^{T},\,i\in\mathbb{N}:\:\\
&\sum_{t=1}^{T}\max_{i\in\mathbb{N}}|c_{i,\,t+1}-c_{i,\,t}|\leq \mathcal{V}_{T} \},\: \forall T\geq2,
\end{split}
\end{align}
where $\{\mathcal{V}_{T}\}_{T\geq1}$ is a sequence of positive numbers. Since $c_{i,\,t}\in[0,\,1/m]$, $\forall i\in\mathbb{N}$ and $t\geq1$, $\max_{i\in\mathbb{N}}|c_{i,\,t+1}-c_{i,\,t}|\ll 1/m$ holds for all $t\geq1$. Thus, $\mathcal{V}_{T}\leq O(T)$ and $\mathcal{V}_{T}/T\ll 1/m$. Meanwhile, $\mathcal{V}_{T}$ is monotonically increasing in $T$. Based on these considerations, the following assumption on $\mathcal{V}_{T}$ is proposed.

%%%%%%%%%%%%%%%%%%%%%%%%%%%%%%%%%%%%%%%%%%%%%%%%%%%%%%%%%%%%%%%%%%%%%%%%%%%%%%%%%%%%%%%%%%%%%%%%%%%%
\begin{assumption}\label{assumption cost}
There exists $T_{0}\geq1$, such that for all $T\geq T_{0}$, $1/m\leq \mathcal{V}_{T}\leq T/m$.
\end{assumption}
The restriction of $T \geq T_{0}$ for some $T_{0}$ in Assumption \ref{assumption cost} is necessary to further present our results. Actually the condition $1/m\leq \mathcal{V}_{T}$ cannot be directly concluded based on the previous discussions. Since $\mathcal{V}_{T}$ is the accumulation of $\max_{i\in\mathbb{N}}|c_{i,\,t+1}-c_{i,\,t}|$ and we suppose that $\max_{i\in\mathbb{N}}|c_{i,\,t+1}-c_{i,\,t}|\ll 1/m$ . Hence, when $T$ is small, $\mathcal{V}_{T}$ may not satisfy the condition $1/m\leq \mathcal{V}_{T}$.
It is worth noting that a justification of the existence of $T_{0}$ is useful to guarantee the performance of our algorithm, while it is not necessary to know the exact value of $T_0$. Prior knowledge can be incorporated to ensure the existence of $T_{0}$. For example, if the difference term $\max_{i\in\mathbb{N}}|c_{i,\,t+1}-c_{i,\,t}|$ fails to rapidly become infinitely small (which is natural when the cost sequences are non-stationary), then $\mathcal{V}_{T}$ will surely be larger than $1/m$ when $T$ is large enough. Examples of securing an estimated value for $T_{0}$ can be found in the smart grid application of Section \ref{section application in smart grid}, where $\mathcal{V}_{T}$ is estimated to grow linearly in $T$ and a threshold time for $\mathcal{V}_{T}$ to exceed $1/m$ can be easily calculated.

%%%%%%%%%%%%%%%%%%%%%%%%%%%%%%%%%%%%%%%%%%%%%%%%%%%%%%%%%%%%%%%%%%%%%%%%%%%%%%%%%%%%%%%%%%%%%%%%%%%%%%
\section{Thompson-Hedge algorithm}\label{section thompson hedge algorithm}
In this section, an online learning algorithm is developed to optimize the Bayesian sup regret for our problem. The algorithm has two layers. In the outer layer, at each time $t\geq1$, the algorithm uses Thompson sampling (posterior sampling) method to sample a parameter vector $\left(\lambda_{1},\cdots,\lambda_{N}\right)$ from the posterior distributions of all $\lambda_{i}$'s based on the historical observations. Then, in the inner layer, a so-called sub-algorithm is fed with the sampled parameters. The sub-algorithm returns an action $\pi_{t}\in\mathbb{N}$. At the end of this loop, the algorithm chooses node $\pi_{t}$, observes the reward from $\pi_{t}$ and updates the posterior distribution of $\lambda_{\pi_{t}}$. In the end of this section, a theorem is provided to characterize the convergence rate of the regret function.

Since Hedge algorithm is applied as the sub-algorithm in the inner layer, the proposed algorithm is named as Thompson-Hedge algorithm. Before formally presenting our algorithm, the Hedge algorithm is firstly introduced. Hedge algorithm is a classical algorithm designed for the adversarial MNB problem with full feedback \citep{freund1999adaptive}. It is a randomized algorithm that maintains a weight $\omega_{t}(i)$ for each $i\in\mathbb{N}$. Hedge algorithm then chooses node $i\in\mathbb{N}$ with a probability proportional to $\omega_{t}(i)$ at time $t$. Subsequently, the weight for each node is updated according to the observed reward of this node. The algorithm chooses a size $\Delta$ and restarts updating the weights of each node every $\Delta$ times. The set of intervals between two successive restarting epochs (including the first restarting time) is called a batch. Since Hedge algorithm is designed for non-Bayesian adversarial MNB, in order to use it under the Bayesian framework, one needs to modify the original Hedge algorithm and feed it by the value of parameter $\lambda$. Denote Hedge($\lambda$) as the modified Hedge algorithm used in our Bayesian adversarial MNB fed by $\lambda$. The algorithm is summarized in Algorithm 1.
%%%%%%%%%%%%%%%%%%%%%%%%%%%%%%%%%%%%%%%%%%%%%%%%%%%%%%%%%%%%%%%%%%%%%%%%%%%%%%%
\begin{algorithm}[h]
\captionsetup{labelformat = empty}
\renewcommand{\algorithmicrequire}{\textbf{Initialization}:}
\caption{\textbf{Algorithm 1: Hedge($\lambda$) Algorithm}}
\label{algorithm hedge}
\begin{algorithmic}[1]
\Require Parameter vector $\lambda$, $\varepsilon=\left(1-\sqrt{\ln{N}/2T}\right)^{-1}$, calculate $\mu_{i}$ using $\lambda_{i}$.
\For{$b=1,2,\cdots,\left\lceil T/\Delta \right\rceil$}
\State For all $i\in\mathbb{N}$, $\omega_{1}^{i}=1$.
\While{$t\leq \min\left\{T,\,b\cdot \Delta\right\}$}
\State For each $i\in\mathbb{N}$, set
	   $$p^{i}_{t}=\frac{\omega_{t}^{i}}{\sum_{j\in\mathbb{N}}\omega_{t}^{j}}.$$
\State Choose a node $\pi_{t}$ according to the probability distribution
       $\left\{p^{i}_{t}\right\}_{i\in\mathbb{N}}$ and receive a reward
       $k_{\pi_{t},\,t}\cdot c_{\pi_{t},\,t}$.
\State For each node $i\in\mathbb{N}$, update
       $$\omega_{t+1}^{i}=\omega_{t}^{i}\varepsilon^{\mu_{i}\cdot c_{\pi_{t},\,t}}.$$
\State Set $t\rightarrow t+1$.
\EndWhile
\EndFor
\end{algorithmic}
\end{algorithm}

Hedge($\lambda$) follows the paradigm of the classical Hedge algorithm. The following lemma shows that if the value of parameter $\lambda$ fed to Hedge($\lambda$) is the true parameter of the Bayesian adversarial MNB model, then Hedge($\lambda$) retains a convergent sup regret with a known upper bound.
\begin{lemma}\label{lemma hedge}
If the input $\lambda$ in the Hedge algorithm is the true model parameter and the batch size is chosen to be $\Delta=\left\lceil\left(\ln{N}\right)^{1/3}\left({T}/{m\mathcal{V}_{T}}\right)^{2/3}\right\rceil$, then for all $T\geq N$, the sup regret by the Hedge($\lambda$) algorithm is upper bounded by
\begin{equation}\label{lemma hedge 1}
\sup_{\vec{c}_{T}\in\mathbb{C}_{T}} R^{\pi}\left(\lambda,\,\vec{c}_{T},\,T\right)\leq
\left(8+2\sqrt{2}\right)\left(m\mathcal{V}_{T}\ln{N}\right)^{1/3}\left(T\right)^{2/3}.
\end{equation}
\end{lemma}
Detailed proof of Lemma \ref{lemma hedge} is given in the Appendix. In the proof of our main result of Theorem \ref{theorem thompson-hedge algorithm}, the conclusion of Lemma \ref{lemma hedge} is used as an intermediate benchmark result.

In the proposed Thompson-Hedge algorithm, parameter $\lambda$ is sampled from its posterior distribution in each epoch. Then, the sampled parameter is fed to the inner algorithm Hedge($\lambda$) and a node is chosen correspondingly. In particular, the algorithm chooses a node according to the probability weight of each node. Then, the algorithm observes the costs and updates the posterior distribution of $\lambda$. Next, a sampled parameter $\hat{\lambda}$ is drawn from the posterior distribution of $\lambda$. Finally, the algorithm updates the weight of each node for next epoch using both the observed costs and $\hat{\lambda}$. $\textup{Gamma}(\alpha,\,\beta)$ is used to represent Gamma distribution with parameters $(\alpha,\,\beta)$. The proposed Thompson-Hedge algorithm is summarized in Algorithm 2.

%%%%%%%%%%%%%%%%%%%%%%%%%%%%%%%%%%%%%%%%%%%%%%%%%%%%%%%%%%%%%%%%%%%%%%%%%%%%%%%%%%
\begin{algorithm}[h]
\captionsetup{labelformat = empty}
\renewcommand{\algorithmicrequire}{\textbf{Initialization}:}
\caption{\textbf{Algorithm 2: Thompson-Hedge Algorithm}}
\label{algorithm thompson-hedge}
\begin{algorithmic}[1]
\Require Parameter $\theta$, batch size $\Delta$, $\varepsilon=\left(1-\sqrt{\ln{N}/2T}\right)^{-1}$, t=1.
\For{$b=1,2,\cdots,\left\lceil T/\Delta \right\rceil$}
\State For all $i\in\mathbb{N}$, $\omega_{1}^{i}=1$.
\While{$t\leq \min\left\{T,\,b\cdot \Delta\right\}$}
\State For each $i\in\mathbb{N}$, set
		       $$p^{i}_{t}=\frac{\omega_{t}^{i}}{\sum_{j\in\mathbb{N}}\omega_{t}^{j}}.$$
\State Choose a node $i_{t}$ according to the probability distribution
       $\left\{p^{i}_{t}\right\}_{i\in\mathbb{N}}$ and receive a reward
       $k_{i_{t},\,t}\cdot c_{i_{t},\,t}$.
\State For node $i_{t}$, update
       $\left(\alpha_{i_{t}},\,\beta_{i_{t}}\right)\rightarrow
       \left(\alpha_{i_{t}}+k_{i_{t},\,t},\,\beta_{i_{t}}+1\right)$.
\State For each node $i\in\mathbb{N}$, sample $\hat{\lambda}_{i}\sim
       \textup{Gamma}(\alpha_{i},\,\beta_{i})$, calculate $\hat{\mu}_{i}$ using $\hat{\lambda}_{i}$, and update $$\omega_{t+1}^{i}=\omega_{t}^{i}\cdot\varepsilon^{\hat{\mu}_{i}\cdot c_{i_{t},\,t}}.$$
\State Set $t\rightarrow t+1$.
\EndWhile
\EndFor
\end{algorithmic}
\end{algorithm}

Before the main result (Theorem \ref{theorem thompson-hedge algorithm}) is presented, the following lemma is introduced that will be used for the proof of the main result.
\begin{lemma}\label{lemma osband}[\citep{osband2013more}, Lemma 2]
If $F$ is the true distribution of $\lambda$, and $\hat{\lambda}$ is the sampled parameter in epoch $t$, then for any $\sigma(\mathbb{H}_{t})$-measurable function $g$, it follows
\begin{equation*}
\mathbf{E}_{\lambda}\left[g(\lambda)\,|\,\mathbb{H}_{t}\right]
=\mathbf{E}_{\hat{\lambda}}\left[g(\hat{\lambda})\,|\,\mathbb{H}_{t}\right].
\end{equation*}
\end{lemma}
Lemma \ref{lemma osband} shows a central result in Bayesian learning area. The sampled parameter from the posterior distribution each time can be ``considered" as the true parameter in the sense that any deterministic function using it as an argument independently has the same expectation. On the basis of Lemma \ref{lemma hedge} and Lemma \ref{lemma osband}, the main theorem is presented as below, which establishes an upper bound on the Bayesian sup regret for our Thompson-Hedge algorithm and thus indicates its convergence rate.

\begin{theorem}\label{theorem thompson-hedge algorithm}
Let $\pi$ be the Thompson-Hedge algorithm with batch size $\Delta=\left\lceil\left(\ln{N}\right)^{1/3}\left({T}/{m\mathcal{V}_{T}}\right)^{2/3}\right\rceil$, $\mathbb{C}_{T}$ be defined in (\ref{equation CT}), and $T_{0}$ be the same value as defined in Assumption \ref{assumption cost}, then for all $T\geq T_{0}$, the Bayesian sup regret $\mathcal{R}^{\pi}\left(\mathbb{C}_{T},\,T\right)$ can be upper bounded by
\begin{equation}\label{equation theorem thompson-hedge algorithm}
\mathcal{R}^{\pi}\left(\mathbb{C}_{T},\,T\right)\leq
\left(8+2\sqrt{2}\right)\left(m\mathcal{V}_{T}\ln{N}\right)^{1/3}\left(T\right)^{2/3}.
\end{equation}
\end{theorem}

\begin{pf}
The basic idea for the proof is sketched as follows. The Bayesian sup regret by Thompson-Hedge algorithm is firstly decomposed into two parts: the regret of the introduced Hedge($\lambda$) algorithm and the difference of performance between Thompson-Hedge and Hedge algorithms. Since the sup regret by Hedge algorithm is bounded by Lemma \ref{lemma hedge},  Theorem \ref{theorem thompson-hedge algorithm} can then be proved by upper bounding the performance difference between the two algorithms. To distinguish between two algorithms in the work, $\pi^{TH}$ and $\pi^{H}$ are used for Thompson-Hedge and Hedge algorithms correspondingly. Moreover, $R^{TH}$ and $R^{H}$ are denoted as the regret function of $\pi^{TH}$ and $\pi^{H}$, respectively.

%%%%%%%%%%%%%%%%%%%%%%%%%%%%%%%%%%%%%%%%%%%%%%%%%%%%%%%%%%%%%%%%%%%%%%%%%%%%%%%%%%%%%
\textit{a) Separation of the target regret $\mathcal{R}^{TH}$.}

In the first step, the Bayesian sup regret $\mathcal{R}^{TH}$ is separated into a combination of two terms. In Step b) and Step c), the upper bound of these two terms are obtained separately and therefore, the upper bound of $\mathcal{R}^{TH}$ can be readily obtained. The separation of $\mathcal{R}^{TH}$ is given as below.
\begin{align}\label{proof of theorem thompson hedge 1}
&\mathcal{R}^{TH}\left(\mathbb{C}_{T},\,T\right)
= \mathbf{E}_{\lambda\sim\mathcal{F}}
\Bigg[\sup_{\vec{c}_{T}\in\mathbb{C}_{T}}\Bigg\{
\sum_{t=1}^{T}\mu_{i_{t}^{*}}\cdot c_{i_{t}^{*},\,t} \nonumber\\
&-\mathbf{E}^{TH}\left(\sum_{t=1}^{T}\mu_{i_{t}}\cdot c_{i_{t},\,t} \,\Big|\, \lambda,\,\vec{c}_{T}\right)
\Bigg\}\Bigg]\nonumber\\
&\leq
\mathbf{E}_{\lambda\sim\mathcal{F}}
\Bigg[\sup_{\vec{c}_{T}\in\mathbb{C}_{T}}\Bigg\{
\sum_{t=1}^{T}\mu_{i_{t}^{*}}\cdot c_{i_{t}^{*},\,t}\nonumber\\
&-\mathbf{E}^{H}\left(\sum_{t=1}^{T}\mu_{i_{t}i}\cdot c_{i_{t},\,t} \,\Big|\, \lambda,\,\vec{c}_{T}\right)
\Bigg\}\Bigg]\nonumber\\
& +\mathbf{E}_{\lambda\sim\mathcal{F}}
\Bigg[\sup_{\vec{c}_{T}\in\mathbb{C}_{T}}\Bigg\{
\mathbf{E}^{H}\left(\sum_{t=1}^{T}\mu_{i_{t}}\cdot c_{i_{t},\,t} \,\Big|\, \lambda,\,\vec{c}_{T}\right)\nonumber\\
&-\mathbf{E}^{TH}\left(\sum_{t=1}^{T}\mu_{i_{t}}\cdot c_{i_{t},\,t} \,\Big|\, \lambda,\,\vec{c}_{T}\right)
\Bigg\}\Bigg].
\end{align}
For notational convenience, denote the above two terms in the two square brackets by $\Lambda_{1}$ and $\Lambda_{2}$, respectively. The relation in (\ref{proof of theorem thompson hedge 1}) can be rewritten as
\begin{equation*}
\mathcal{R}^{TH} = \Lambda_{1} + \Lambda_{2}
\end{equation*}
Note that $\Lambda_{1}$ is closely associated with the sup regret of $\pi^{H}$ and can be bounded based on the result in Lemma \ref{lemma hedge}. $\Lambda_{2}$ is the difference between the return of $\pi^{TH}$ and $\pi^{H}$ that will be bounded in the following step.

%%%%%%%%%%%%%%%%%%%%%%%%%%%%%%%%%%%%%%%%%%%%%%%%%%%%%%%%%%%%%%%%%%%%%%%%%%%%%%%%%%%%%
\textit{b) Upper bound of $\Lambda_{1}$}

$\Lambda_{1}$ is rewritten as
\begin{align*}
\Lambda_{1}=\mathbf{E}_{\lambda\sim\mathcal{F}}
\left[\sup_{\vec{c}_{T}\in\mathbb{C}_{T}} R^{H}\left(\lambda,\,\vec{c}_{T},\,T\right)\right].
\end{align*}
Note that the upper bound in Lemma \ref{lemma hedge} holds for any true model parameter $\lambda$. Thus the upper bound still holds after taking expectation on $\lambda$ on both sides of (\ref{lemma hedge 1}). Therefore, it holds that
\begin{align*}
\Lambda_{1}\leq
\left(8+2\sqrt{2}\right)\left(m\mathcal{V}_{T}\ln{N}\right)^{1/3}\left(T\right)^{2/3}.
\end{align*}
%

%%%%%%%%%%%%%%%%%%%%%%%%%%%%%%%%%%%%%%%%%%%%%%%%%%%%%%%%%%%%%%%%%%%%%%%%%%%%%%%%%%%%%
\textit{c) Upper bound $\Lambda_{2}$}

 The following clarification is made for notational convenience. $\lambda_{t}$ is used for the sampled parameter by Thompson-Hedge algorithm at time $t$ and $\lambda$ for the true model parameter which is the input of Hedge algorithm. To bound the Bayesian sup regret by Thompson-Hedge algorithm, the difference between the Bayesian sup regret functions by Thompson-Hedge and Hedge algorithms, respectively, is firstly bounded, namely,
\begin{align}\label{equation proof theorem thompson-hedge 1}
\begin{split}
\mathbf{E}_{\lambda\sim\mathcal{F}}
\Bigg[\sup_{\vec{c}_{T}\in\mathbb{C}_{T}} \Bigg\{
&\mathbf{E}^{TH}\left(\sum_{t=1}^{T}\mu_{i_{t}}\cdot c_{i_{t},\,t} \,\Big|\,\lambda,\,\vec{c}_{T}\right)\\
&-\mathbf{E}^{H}\left(\sum_{t=1}^{T}\mu_{i_{t}}\cdot c_{i_{t},\,t} \,\Big|\,\lambda,\,\vec{c}_{T}\right)\Bigg\}\Bigg],
\end{split}
\end{align}
where $\mathbf{E}^{TH}$ and $\mathbf{E}^{H}$ have the same meaning as $\mathbf{E}^{\pi}$ given $\pi$, for Thompson-Hedge algorithm and Hedge algorithm, respectively. Denote the observation history before time $t\geq2$ as $\mathbb{H}_{t}=\left\{i_{1},\,k_{i_{1},\,1},\,c_{1},\cdots,i_{t-1},\,k_{i_{t-1},\,t-1},\,c_{t-1}\right\}$. Conditioned on the observation history, the probability weight function $p_{t}(\cdot)$ in both algorithms is functions of $\lambda$ and $\lambda(t)$ correspondingly. Moreover, let $p_{t}(\cdot)$ and $\Tilde{p}_{t}^{i}(\cdot)$ be the probability functions of Thompson-Hedge and Hedge algorithms, respectively. For any fixed $\vec{c}_{T}\in\mathbb{C}_{T}$, the following relation holds,
\begin{align}\label{equation proof theorem thompson-hedge 2}
& \mathbf{E}^{TH}\left(\sum_{t=1}^{T}\mu_{i_{t}}\cdot c_{i_{t},\,t} \,\Big|\,\lambda,\,\vec{c}_{T}\right)\nonumber\\
&-\mathbf{E}^{H}\left(\sum_{t=1}^{T}\mu_{i_{t}}\cdot c_{i_{t},\,t} \,\Big|\,\lambda,\,\vec{c}_{T}\right)\nonumber\\
= &
\mathbf{E}^{TH}\left(\sum_{t=1}^{T}\mu_{i_{t}}\cdot c_{i_{t},\,t} \,\Big|\,\lambda,\,\vec{c}_{T},\,\mathbb{H}_{T}\right)\nonumber\\
&-\mathbf{E}^{H}\left(\sum_{t=1}^{T}\mu_{i_{t}}\cdot c_{i_{t},\,t} \,\Big|\,\lambda,\,\vec{c}_{T},\,\mathbb{H}_{T}\right)\nonumber\\
= &
\sum_{t=1}^{T}\mathbf{E}_{\lambda_{t}}\left[p_{t}^{i}(\lambda_{t})-
\Tilde{p}_{t}^{i}(\lambda)|\mathbb{H}_{t}\right]\mu_{i}\cdot c_{i,\,t}.
\end{align}
At any time $t\geq1$, note that $\lambda_{t}$ is the sampled parameter from the same posterior distribution as the true $\lambda$. Meanwhile, $p_{t}^{i}(\cdot)$ and $\Tilde{p}_{t}^{i}(\cdot)$ are the same deterministic function based on $\mathbb{H}_{t}$. According to Lemma \ref{lemma osband}, it follows
\begin{equation*}
\mathbf{E}_{\lambda_{t},\,\lambda}\left[p_{t}^{i}(\lambda_{t})-
\Tilde{p}_{t}^{i}(\lambda)|\mathbb{H}_{t}\right]\mu_{i}\cdot c_{i,\,t}=0.
\end{equation*}
Therefore, for any fixed $\vec{c}_{T}\in\mathbb{C}_{T}$, it holds that
\begin{align*}
&\mathbf{E}_{\lambda\sim\mathcal{F}}\Bigg[\Bigg\{
\mathbf{E}^{TH}\left(\sum_{t=1}^{T}\mu_{i_{t}}\cdot c_{i_{t},\,t} \,\Big|\,\lambda,\,\vec{c}_{T}\right)\\
&-\mathbf{E}^{H}\left(\sum_{t=1}^{T}\mu_{i_{t}}\cdot c_{i_{t},\,t} \,\Big|\,\lambda,\,\vec{c}_{T}\right)
\Bigg\}\Bigg]\nonumber\\
 &=\mathbf{E}\Bigg(\mathbf{E}_{\lambda\sim\mathcal{F}}
\Bigg[\Bigg\{
\mathbf{E}^{TH}\left(\sum_{t=1}^{T}\mu_{i_{t}}\cdot c_{i_{t},\,t} \,\Big|\,\lambda,\,\vec{c}_{T},\,\mathbb{H}_{T}\right)\\
&-\mathbf{E}^{H}\left(\sum_{t=1}^{T}\mu_{i_{t}}\cdot c_{i_{t},\,t} \,\Big|\,\lambda,\,\vec{c}_{T},\,\mathbb{H}_{T}\right)
\Bigg\}\Bigg]\Bigg)\nonumber\\
&=
\mathbf{E}\left(\mathbf{E}_{\lambda_{t},\,\lambda}\left[p_{t}^{i}(\lambda_{t})-
\Tilde{p}_{t}^{i}(\lambda)|\mathbb{H}_{t}\right]\mu_{i}\cdot c_{i,\,t}\right)\nonumber\\
& = 0.
\end{align*}
Note that the relation above holds for any $\vec{c}_{T}\in\mathbb{C}_{T}$, which leads to
\begin{align}
\begin{split}
\mathbf{E}_{\lambda\sim\mathcal{F}}
\Bigg[\sup_{\vec{c}_{T}\in\mathbb{C}_{T}} \Bigg\{
&\mathbf{E}^{TH}\left(\sum_{t=1}^{T}\mu_{i_{t}}\cdot c_{i_{t},\,t} \,\Big|\,\lambda,\,\vec{c}_{T}\right)\\
&-\mathbf{E}^{H}\left(\sum_{t=1}^{T}\mu_{i_{t}}\cdot c_{i_{t},\,t} \,\Big|\,\lambda,\,\vec{c}_{T}\right)\Bigg\}\Bigg]\equiv 0.
\end{split}
\end{align}
%

%%%%%%%%%%%%%%%%%%%%%%%%%%%%%%%%%%%%%%%%%%%%%%%%%%%%%%
\textit{d) Upper bound of the regret $\mathcal{R}^{TH}$}.

Finally, the upper bound of the Bayesian sup regret of Thompson-Hedge algorithm is obtained by combining the results in a), b) and c). In particular, it holds that
\begin{align*}
\mathcal{R}^{TH}\left(\mathbb{C}_{T},\,T\right)
&\leq \Lambda_{1} + \Lambda_{2} = \Lambda_{1} + 0 \nonumber\\
&\leq \left(8+2\sqrt{2}\right)\left(m\mathcal{V}_{T}\ln{N}\right)^{1/3}\left(T\right)^{2/3}.
\end{align*}
Therefore, the proof is concluded.
\end{pf}

\begin{remark}\label{discussion of theorem1}
Relevant research that considers a similar problem can be found in \citet{besbes2019optimal}. In \citet{besbes2019optimal}, the classical EXP3 type algorithm was used and an upper bound of the order $O\left(\left(m\mathcal{V}_{T}N\ln{N}\right)^{1/3}\left(T\right)^{2/3}\right)$ was obtained for the sup regret. Since the upper bound holds uniformly on the parameter space, the same upper bound also holds for the Bayesian sup regret. If the constant $\left(8+2\sqrt{2}\right)$ in (\ref{equation theorem thompson-hedge algorithm}) is neglected, our bound outperforms the bound in \citet{besbes2019optimal} by a term of $O(N^{1/3})$, which implies that the performance of our Thompson-Hedge algorithm is less sensitive to the number of nodes $N$. It indicates that when considering problems with large scales, our proposed algorithm is supposed to retain a Bayesian sup regret that converges relatively faster. Meanwhile, \citet{besbes2019optimal} constructs a lower bound of the order $O\left(\mathcal{V}_{T}^{1/3}T^{2/3}\right)$ on the regret by any algorithm. Similarly, the lower bound also holds uniformly on all the parameters and adapts to our problem. Therefore, the lower bound shows that our algorithm achieves the order optimality.
\end{remark}

%%%%%%%%%%%%%%%%%%%%%%%%%%%%%%%%%%%%%%%%%%%%%%%%%%%%%%%%%%%%%%%%%%%%%%%%%%%%%%%%%%%%%%%%%%%%%%%%%%
\section{Applicability in smart grids}\label{section application in smart grid}
 The model is formulated following the learning context in Section \ref{section problem formulation} and an online learning algorithm is developed in Section \ref{section thompson hedge algorithm}. This section presents an application case to demonstrate the applicability of the proposed model and method in practical smart grids. In particular, Assumption \ref{assumption cost} plays a central role in our model. A real data set is used to verify that Assumption \ref{assumption cost} may hold in reality, and thus the proposed method is practical. Section \ref{subsection operation cost of the smart grid} introduces the procedure for calculating the reward $c_{i,\,t}$ in the smart grid. In Section \ref{subsection linear regression}, linear regression method is used to show a linear growth rate of the critical quantity $\mathcal{V}_{T}$ in Assumption \ref{assumption cost}, and thus it is concluded that Assumption \ref{assumption cost} holds for the selected data set.

%%%%%%%%%%%%%%%%%%%%%%%%%%%%%%%%%%%%%%%%%%%%%%%%%%%%%%%%%%%%%%%%%%%%%%%%%%%%%%%%%%%%%%%%%%%%%%%%%%%%
\subsection{Operation cost of the smart grid}\label{subsection operation cost of the smart grid}

To facilitate reading, meanings of the variables used in calculating the operation cost of the smart grid, are displayed in different categories as follows. For the operational variables, $\vartheta _t$ is the operation state of the smart grid at time $t$, $P_{i,j}^t$ [kW] is the output of type $j$ power source at Node $i$, $\bar P _{i,j}$ is the rate power of type $j$ power source at Node $i$, $L_i^t$ is the power load at Node $i$, $\bar L_i^t$ is the load shedding, $i.e.$, power demand not supplied, at Node $i$, $\varphi$ is the state of charge of the energy storage system (ESS) and $\bar C_{ESS}$ is the power capacity of the ESS. For the configuration parameters of the smart grid, $B_{i,i^{'}}$ [$1/\Omega$] is the susceptance of the pairs of Nodes $(i,i^{'})$, $\delta_i$ is the voltage angle at Node $i$, $\Delta \delta$ is the voltage angle difference $(\delta_i-\delta_{i^{'}})$ of two nearby nodes, $A_{i,i^{'} }$ [A] is the ampacity of the pairs of Nodes $(i,i^{'})$ and $V$ [kW] is the nominal voltage of the smart grid. For the cost and price coefficients, $Co^{\vartheta_t }$ is the operation cost of the smart grid subject to operation state $\vartheta_t$, $Cs_j$ is the variable operation cost for power source $j$, $Cf_{i,i^{'}}$ is the variable operation cost for feeder $(i,i^{'})$, $Cp$ is the penalty cost for power demand not supplied, and $Ep^{\vartheta_t}$ is the energy price associated with operation state $\vartheta_t$. $T_s$ is the duration of $\vartheta_t$, $\mathbb{P}$ denotes a subset of power sources, $\mathbb{N}$ denotes the set of all nodes, and $\mathbb{F}$ denotes the set of node pairs with transmission line between them. $s_{i,i^{'}}=1$ indicates that power cannot be transmitted between Node $i$ and Node $i^{'}$ due to the successful cyber attack on Node $i$.

The linear Direct Current (DC) power flow model \citep{mo2019impact,sahraei2016computationally} is introduced to interprete the physical meaning of $c_{i,\,t}$ in the smart grid. At time $t$, operation state of the distributed generation system (DGS) is denoted by the following vector:

\begin{equation}\label{equation DGS}
\vartheta_{t}=\left[P_{i,\,j}^{t}\,L_{i}^{t}\right],
\end{equation}
where the power sources considered in this work consists of natural gas plant, biomass plant, wind farm, Photovoltaic farm and ESS. These data are all sampled from the historical database provided by the Elia Grid, Belgium\footnote{(URL: https://www.elia.be/en/grid-data)}.

Before calculating $c_{i,\,t}$, the first objective is to achieve the minimal cost in the presence of load shedding, denoted by $Co^{\vartheta_{t}}$, by solving the following linear optimization problem

\begin{align}\label{equation cost objective function}
\min Co^{\vartheta_{t}}
= &
\sum_{i\in \mathbb{N}}\sum_{j\in \mathbb{P}}
\left(Cs_{j}-Ep^{\vartheta_{t}}\right)P_{i,\,j}^{t}\nonumber\\
&
+\sum_{(i,\,i^{'})\in \mathbb{F}}Cf_{i,\,i^{'}}\left|B_{i,\,i^{'}}\left(\delta_{i}-\delta_{i^{'}}\right)\right| \nonumber\\
&+\left(Cp+Ep^{\vartheta_{t}}\right)\sum_{i=1}^{N}\bar L_{i}^{t}
\end{align}
subject to
\begin{equation}\label{equation cost constraint 1}
L_{i}^{t} - \bar L_{i}^{t}-\sum_{j\in \mathbb{P}}P_{i,\,j}^{t} - \sum_{i=1}^{N}B_{i,\,i^{'}}(\delta_{i}-\delta_{i^{'}})=0,
\end{equation}
\begin{equation}\label{equation cost constraint 2}
0\leq P_{i,\,j}\leq \bar P_{i,\,j}^{t},\: j\neq ESS,
\end{equation}
\begin{equation}\label{equation cost constraint 3}
\left|P_{i,\,j}^{t}\cdot T_{s}\right| \leq
\min\left\{(1-\varphi)\cdot\bar C_{ESS},\,\bar P_{i}^{j}\cdot T_{s}\right\},\: j=ESS,
\end{equation}
\begin{equation}\label{equation cost constraint 4}
\left| B_{i,\,i^{'}}(\delta_{i}-\delta_{i^{'}}) \right|\leq (1-s_{i,\,i^{'}})V A_{i,\,i^{'}},
\end{equation}

where constraint (\ref{equation cost constraint 1}) requires that the power generated and consumed is balancing at each node of DGS, constraint (\ref{equation cost constraint 2}) requires that the power generated should not be larger than the rated power, constraint (\ref{equation cost constraint 3}) indicates that the charging or discharging of ESS should not be larger than the remaining capacity or nominal rate, and constraint (\ref{equation cost constraint 4}) indicates that the power flow between two nodes should not be larger than the capacity of the transmission line. Eq. (\ref{equation cost objective function}) is the objective function of a typical linear DC Optimal Power Flow model, which aims to minimize the total operation cost including the cost of generating power, the cost of running feeders and the penalty cost of demand not supplied, when satisfying physical constraints (\ref{equation cost constraint 1})-(\ref{equation cost constraint 4}) of DGS.

The linear DC optimal power flow model is configurated in the Matlab and can be solved using the well-known Simplex method, where the values of configuration parameters have been given in \citet{mo2019impact} and the computation time of operation cost for each practical operation state is around 0.006 second in Gurobi. This computation time contributes most to the total simulation time and therefore our proposed algorithm can be implemented in real time.

The physical meaning of $c_{i,\,t}$ in the smart grid can be defined as the difference between the operation cost of the DGS without cyber attacks and the operation cost of the DGS given the Node $i^{*}$ is temporarily unavailable caused by a successful cyber attack. Therefore, when probing Node $i$ at time $t$, the reward function $c_{i,\,i}$ is defined as

\begin{equation}\label{equation final cost}
\begin{split}
c_{i,\,t}\triangleq &\Big|Co^{\vartheta_{t}}\left(P_{i,j}^t,\,L_i^{t},\,\Delta\delta;\,s_{i,\,i^{'}}=0\right)\\
&-Co^{\vartheta_{t}}\left(P_{i,j}^t,\,L_i^{t},\,\Delta\delta,\,s_{i,\,i^{'}}\,;\,s_{i,\,i^{'}}=1\right)\Big|,
\end{split}
\end{equation}
where $Co^{\vartheta_{t}}\left(P_{i,j}^t,\,L_i^{t},\,\Delta\delta;\,s_{i,\,i^{'}}=0\right)$ is the operation cost of the DGS without cyber attacks, and $Co^{\vartheta_{t}}\left(P_{i,j}^t,\,L_i^{t},\,\Delta\delta;\,s_{i,\,i^{'}}=1\right)$ is the operation cost of the DGS given Node $i$ is unavailable caused by the successful cyber attack on Node $i$. That is to say, the reward $c_{i,t}$ can be calculated via solving the linear optimization problem, defined by (\ref{equation cost objective function}) through (\ref{equation cost constraint 4}) (with/without cyber attacks), where the input $\vartheta_t$ is drawn from the dataset of Elia Grid, Belgium.

%%%%%%%%%%%%%%%%%%%%%%%%%%%%%%%%%%%%%%%%%%%%%%%%%%%%%%%%%%%%%%%%%%%%%%%%%%%%%%%%%%%%%%%%%%%%%%%
\subsection{Numerical analysis of cost sequences}\label{subsection linear regression}
\begin{figure}
\centering
\includegraphics[width=0.4\textwidth]{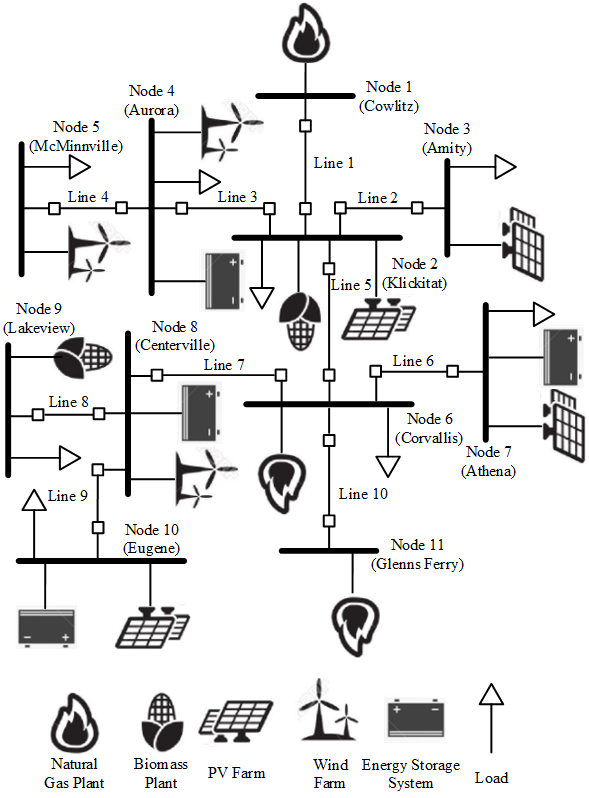}
\caption{The $11$-node radial DGS originated from the IEEE 13 node test feeder.}
\label{figure The 11-node radial DGS originated from the IEEE 13 node test feeder}
\end{figure}

\begin{figure}
\centering
\includegraphics[width=0.46\textwidth]{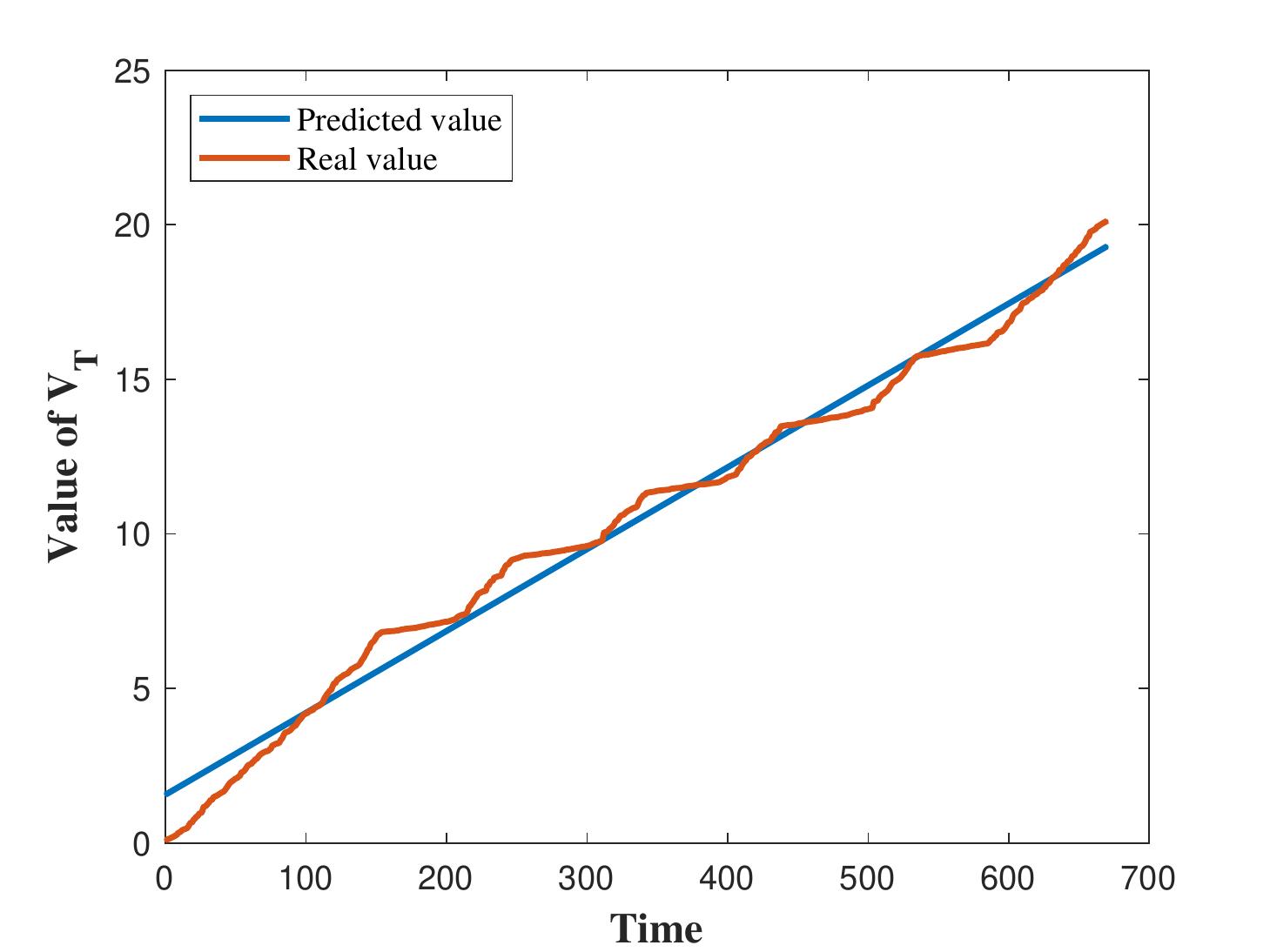}
\caption{Data scatter plot and linear prediction of $\mathcal{V}_{T}$.}
\label{figure the scatter plot}
\end{figure}

In this section, numerical analysis is presented based on a real data set to verify that Assumption \ref{assumption cost} holds in reality. Note that Assumption \ref{assumption cost} implies a linear or sub-linear upper bound in terms of $T$ on $\mathcal{V}_{T}$. Therefore, if $\mathcal{V}_{T}$ has a linear or sub-linear growth rate in $T$, then Assumption \ref{assumption cost} is supposed to hold by choosing a proper value for $m$. Thus, linear regression is performed on the sequence of $\mathcal{V}_{T}$ against time $T$. The realistic grid data from the Elia is used, which provides data from the Belgian electricity market system. The underlying electricity network (a subgrid of Elia Grid) is shown in Figure \ref{figure The 11-node radial DGS originated from the IEEE 13 node test feeder}. In the DGS, the dataset of Elia Grid, Belgium is recorded every 15 minutes, which means that the optimal power flow model will be performed and generate one attack cost $c_{i,t}$ every 15 minutes. The DGS under cyber attacks is investigated over one week, which indicates that a specific dataset with 672 successive observations of the attack cost is selected for illustration. $m$ is chosen to be $4$ and the data has been normalized such that $c_{i,\,t}\in[0,\,1/4]$. The scatter plot is presented in Figure \ref{figure the scatter plot}.

From Figure \ref{figure the scatter plot}, a significant linear relation between $\mathcal{V}_{T}$ and $T$ can be observed. For further justification, the numerical results of the linear regression are presented in Table \ref{table coefficients}.
\begin{table}[h]
    \centering
     \caption{Regression results.}
    \setlength{\tabcolsep}{4pt}

    \begin{tabular}{ccccc}
    \hline
     \, & Coefficients & Standard error & t Stat & P-value \\
     \hline
     Intercept & 1.560 & 0.04711 & 32.04 & $<0.01$\\
     $T$ & 0.02648 & 0.0001212 & 218.5 & 0 \\
    \hline
    \end{tabular}
    \label{table coefficients}
\end{table}

According to the results in Table \ref{table coefficients}, the coefficient for $T$ is about $0.03$, which means that Assumption 1 holds with $T_{0}$ chosen to be $9$. As a significance test for the regression model, results of ANOVA for the regression model is shown in Table \ref{table ANOVA}.
\begin{table}[h]
    \centering
    \setlength{\tabcolsep}{4pt}
      \caption{ANOVA results.}
    \begin{tabular}{cccccc}
    \hline
     \, & df & SS & MS & F & Significance \\
     \hline
     Regression & 1 & 17655 & 17655 & 47734 & $\le$0.01 \\
     Residual & 669 & 247.4 & 0.3699 & \, & \, \\
     Total & 670 & 17902.4 & \, & \, & \, \\
    \hline
    \end{tabular}

    \label{table ANOVA}
\end{table}

\noindent From the results in Table \ref{table ANOVA}, the significance level is below $0.01$, which means that the linear relation between $\mathcal{V}_{T}$ and $T$ for this data set is significant and the model is justified.

%

%%%%%%%%%%%%%%%%%%%%%%%%%%%%%%%%%%%%%%%%%%%%%%%%%%%%%%%%%%%%%%%%%%%%%%%%%%%%%%%%%%%%%%%%%%%%%%%%%%%%%%%%%%

\section{Numerical results}\label{section simulation study}
In this section, the performance of our Thompson-Hedge algorithm is illustrated through simulation studies. In Section \ref{subsection comparative study}, a comparative study is conducted between the performances of our Thompson-Hedge algorithm and the R.EXP3 algorithm which was recently proposed in \citet{besbes2019optimal}. Subsequently, in Section \ref{subsection sensitivity to the variation},  sensitivity analysis is conducted to investigate the influence of constraint $\mathcal{V}_{T}$ on the cost sequence.

%%%%%%%%%%%%%%%%%%%%%%%%%%%%%%%%%%%%%%%%%%%%%%%%%%%%%%%%%%%%%%%%%%%%%%%%%%%%%%%%%%%%%%%%%%%%%%%%%%%%
\subsection{Comparative study}\label{subsection comparative study}
This section compares the performance of the proposed Thompson-Hedge algorithm with the R.EXP3 algorithm that was also designed for the same adversarial problem with constrained variation on cost sequence. Both algorithms use batch methods from the original EXP3 algorithm, $i.e.$, the time horizon is divided into small batches and the algorithm ``restarts'' at the beginning of each batch. Meanwhile, the R.EXP3 algorithm is also a randomized algorithm, retains a weight function for each node and updates the weights each time. Different from our Thompson-Hedge algorithm, in the R.EXP3 algorithm, if node $i\in\mathbb{N}$ is chosen at any time, the total reward $K_{i,\,t}\cdot c_{i,\,t}$ is considered as a whole adversarial reward. Since the attack number $k_{i^{'},\,t}$ is unknown if $i\neq i^{'}$, the R.EXP3 is supposed to ignore $c_{i^{'},\,t}$ and treat the problem as a typical MNB model with bandit feedback.

The simulation is executed under $N=10$ and $N=20$. To initialize, the parameter $\theta=\left(\alpha_{1},\,\beta_{1},\cdots,\alpha_{N},\,\beta_{N}\right)$ is set for all $i\in\mathbb{N}$, and two large numbers for $Q$ and $L$. The preset parameter $\theta$ is used to generate $\lambda=(\lambda_{1},\cdots,\lambda_{N})$ for $Q$ times in total. Under each generated $\lambda$, $\lambda$ is used to generate the sequence $\{K_{i,\,t}\}_{t=1}^{T}$, and the adversarial cost sequence $\{c_{i,\,t}\}_{t=1}^{T}$ is artificially generated. Then, both algorithms are run based on $\{K_{i,\,t}\}_{t=1}^{T}$ and $\{c_{i,\,t}\}_{t=1}^{T}$, $\forall i\in\mathbb{N}$. Finally, the Bayesian sup regret in the two algorithms are calculated and compared. The simulation process is summarized using pseudo codes in Algorithm 3. Experience from the previous work \citet{smith2018cyber} and the real database from our numerical example are integrated to determine the parameter vector $\theta$. In \citet{smith2018cyber}, the prior distribution is set to be Gamma$(2,\,2)$; in our numerical example. It is estimated that $\lambda\approx 0.25$, which may correspond to the prior distribution Gamma$(1,\,4)$. Therefore, two values for $\theta$ are selected to implement the simulation:
$\theta_{1}:\:\left(\alpha_{1},\,\beta_{1}\right)=\cdots=\left(\alpha_{N},\,\beta_{N}\right)=(2,\,2)$ and
$\theta_{2}:\:\left(\alpha_{1},\,\beta_{1}\right)=\cdots=\left(\alpha_{N},\,\beta_{N}\right)=(1,\,4)$. According to the choices of $\theta$, the truncation parameter $m$ is set as $3$, which ensures that the probability $Pr(K_{i,\,t}>m)$ is small.

%%%%%%%%%%%%%%%%%%%%%%%%%%%%%%%%%%%%%%%%%%%%%%%%%%%%%%%%%%%%%%%%%%%%%%%%%%%
\begin{algorithm}[h]
\captionsetup{labelformat = empty}
\renewcommand{\algorithmicrequire}{\textbf{Initialization}:}
\caption{\textbf{Algorithm 3: Simulation for comparison}}
\label{simulation for comparison}
\begin{algorithmic}[1]
\Require Preset parameter $\theta=(\alpha_{1},\,\beta_{1},\cdots,\alpha_{N},\,\beta_{N})$; numbers of simulation trials $Q$, $L$; time horizon $T$.
\For{$q=1:Q$}
\State For all $i\in\mathbb{N}$, generate $\lambda_{i}$ from $Gamma(\alpha_{i},\,\beta_{i})$.
\For{$l=1:L$}
\For{$t=1:T$}
\For{$i\in\mathbb{N}$}
\State Generate $K_{i,\,t}$ from $Poisson(\lambda_{i})$.
\State Generate $c_{i,\,t}$ artificially.
\EndFor
\EndFor
\State Run Thompson-Hedge algorithm and R.EXP3 algorithm independently based on
       the sequences $\{K_{i,\,t}\}_{t=1}^{T}$ and $\{c_{i,\,t}\}_{t=1}^{T}$, $\forall i\in\mathbb{N}$.
\State Calculate the regret function $R_{l}^{TH}\left(\lambda,\,\vec{c}_{T},\,T\right)$ for Thompson-Hedge
       algorithm and $R_{l}^{R3}\left(\lambda,\,\vec{c}_{T},\,T\right)$ for R.EXP3 algorithm.
\EndFor
\State Calculate the sup regret $\Tilde{R}_{q}^{TH}=\max_{l=1:L}R_{l}^{TH}$ for Thompson-Hedge algorithm and
       $\Tilde{R}_{q}^{R3}=\max_{l=1:L}R_{l}^{R3}$ for R.EXP3 algorithm.
\EndFor
\State Calculate the Bayesian sup regret
       $$\mathcal{R}^{TH}\left(\mathbb{C}_{T},\,T\right)=1/Q\sum_{q=1}^{Q}\Tilde{R}_{q}^{TH},$$
       for Thompson-Hedge algorithm and,
       $$\mathcal{R}^{R3}\left(\mathbb{C}_{T},\,T\right)=1/Q\sum_{q=1}^{Q}\Tilde{R}_{q}^{R3},$$
       for R.EXP3 algorithm.
\end{algorithmic}
\end{algorithm}

Note that in the simulation, the empirical estimation of the sup regret and Bayesian sup regret functions is used in the two algorithms. Therefore, it is necessary to set the values of $Q$ and $L$ large enough to make the estimation with good precision. In the simulation, it is set as $Q=L=10^{5}$. To simulate the real situations, the procedures of sketching the cost sequence are given as below:
 \\

\noindent\textit{Step 1}: For all $i\in\mathbb{N}$, generate $c_{i,\,1}$ uniformly on $(0,\,1/m)$ independently.
 \\

\noindent\textit{Step 2}: For $t=2,\cdots,T$, generate $c_{i,\,t}$ independently from the uniform distribution on the overlapping interval between $(c_{i,\,t-1}-1/100m,\,c_{i,\,t-1}+1/100m)$ and $(0,\,1/m)$.
 \\

\begin{figure}
\centering
\includegraphics[width=0.45\textwidth]{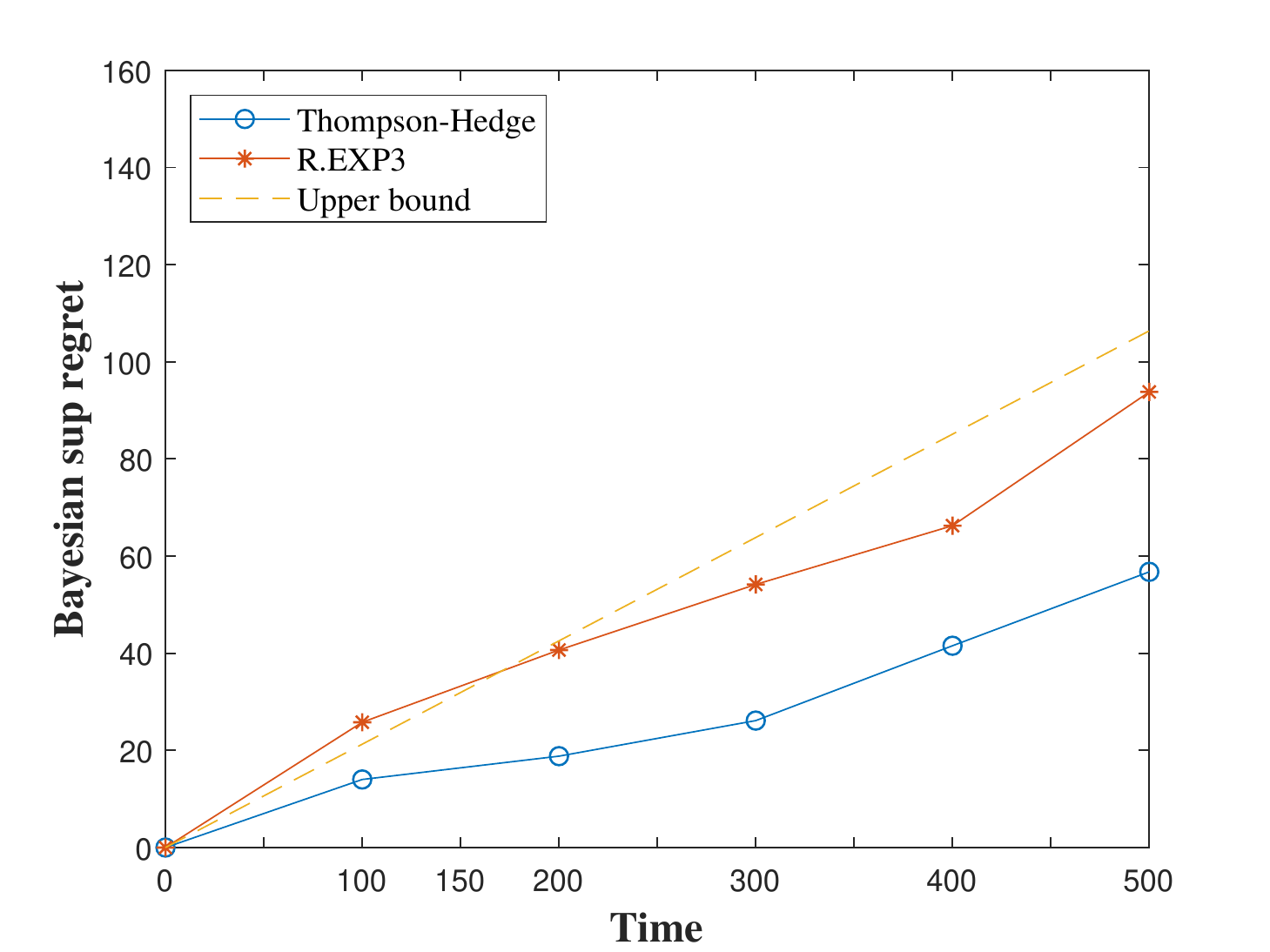}
\caption{Comparison of regrets ($\theta=\theta_{1}$, $N=10$).}
\label{fig11}
\end{figure}

\begin{figure}
\centering
\includegraphics[width=0.45\textwidth]{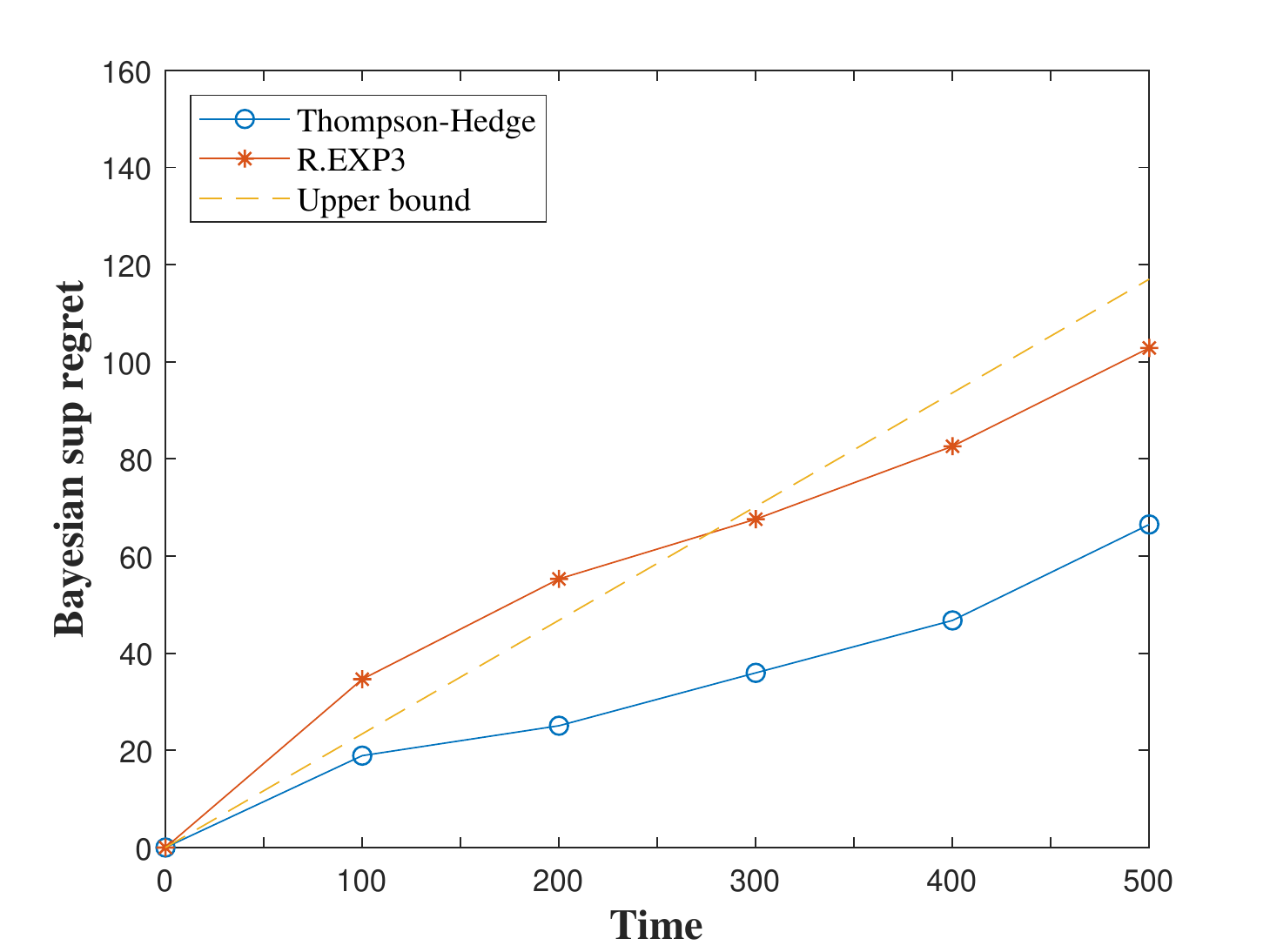}
\caption{Comparison of regrets ($\theta=\theta_{1}$, $N=20$).}
\label{fig12}
\end{figure}

By \textit{Step 1}, simulate the random initial value of each cost sequence. Then, by \textit{Step 2}, set $|c_{i,\,t}-c_{i,\,t-1}|\leq 1/50m$. It can be verified that any cost sequence generated by \textit{Step 1} and \textit{Step 2} satisfies Assumption \ref{assumption cost} with $T_{0}=50$.
The simulation results under $\left(\alpha_{1},\,\beta_{1}\right)=\cdots=\left(\alpha_{N},\,\beta_{N}\right)=(2,\,2)$ are given in
Figures \ref{fig11} and \ref{fig12}.

\begin{figure}
\centering
\includegraphics[width=0.45\textwidth]{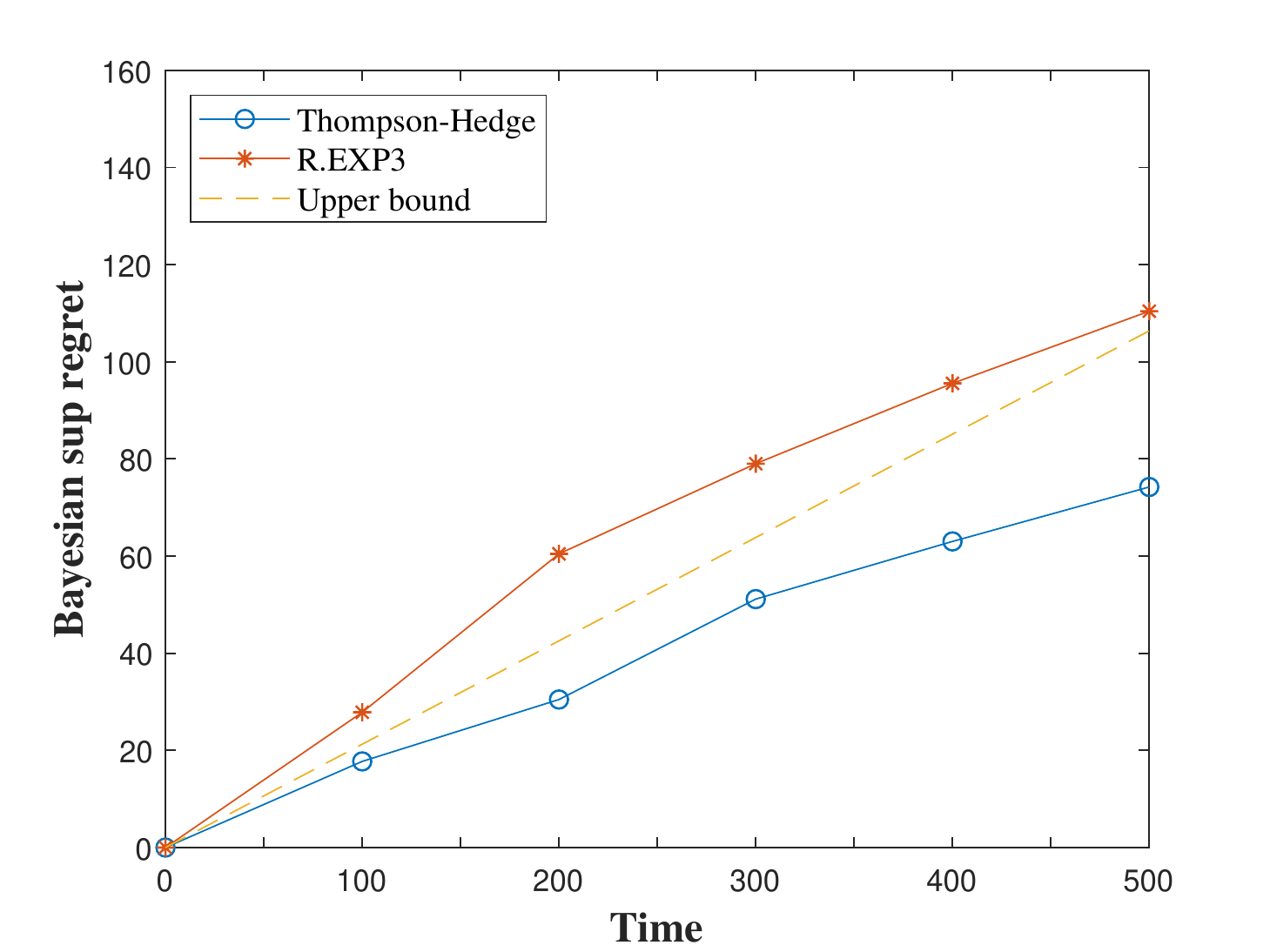}
\caption{Comparison of regrets ($\theta=\theta_{2}$, $N=10$).}
\label{fig21}
\end{figure}

\begin{figure}
\centering
\includegraphics[width=0.45\textwidth]{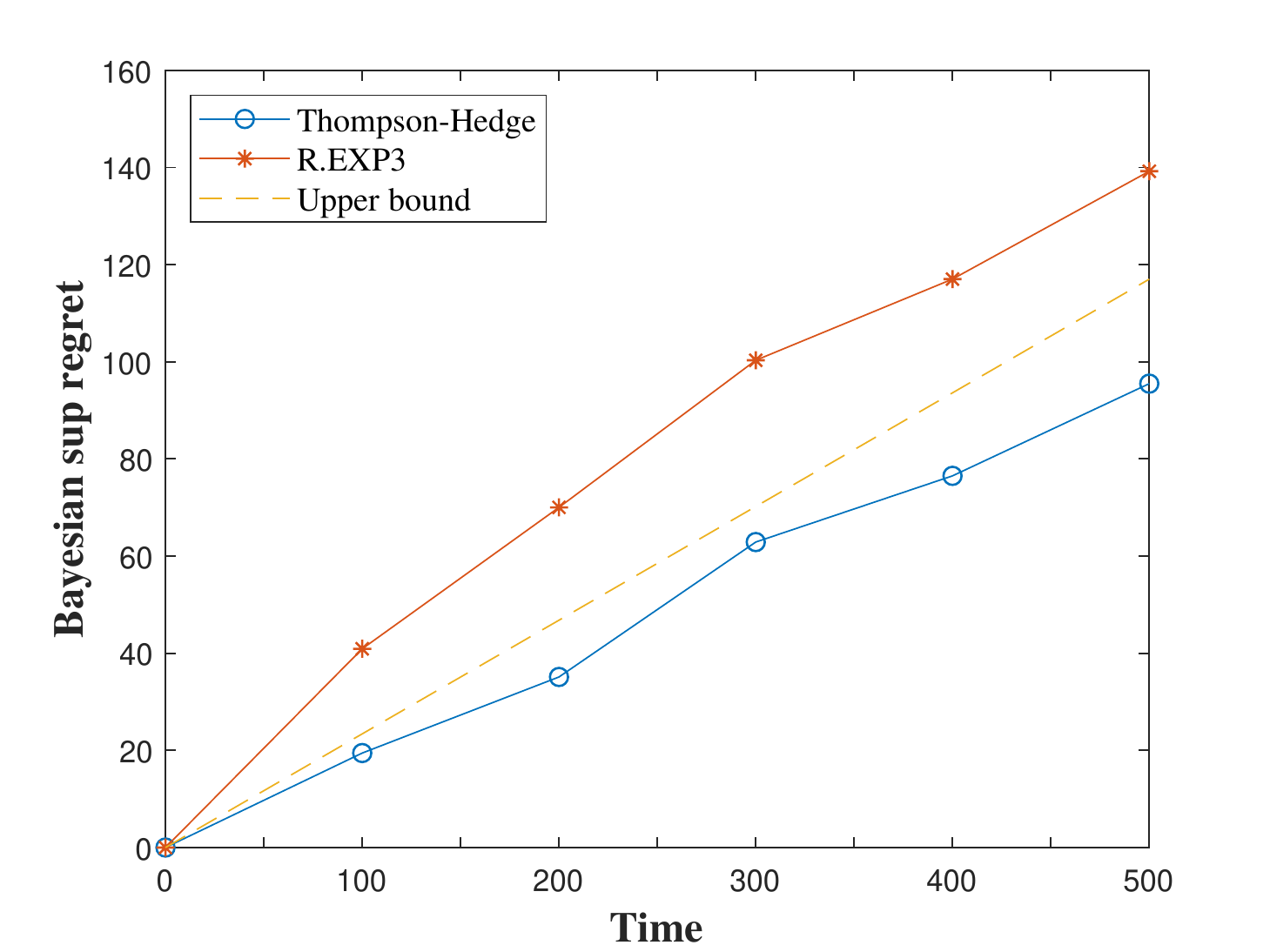}
\caption{Comparison of regrets ($\theta=\theta_{2}$, $N=20$).}
\label{fig22}
\end{figure}

Using the same process, simulation is implemented for $\left(\alpha_{1},\,\beta_{1}\right)=\cdots=\left(\alpha_{N},\,\beta_{N}\right)=(1,\,4)$. The results are given in Figures \ref{fig21} and \ref{fig22}.

In Figures \ref{fig11}-\ref{fig22}, our Thompson-Hedge algorithm outperforms the existing R.EXP3 algorithm in terms of Bayesian sup regret. In particular, Thompson-Hedge algorithm is less sensitive to the problem scale $N$ by comparing the regret curves under $N=10$ and $N=20$, which is consistent with our discussion in Remark \ref{discussion of theorem1}. Our algorithm has advantages over a typical algorithm designed for MNB with bandit feedback, such as the EXP3 or R.EXP3 algorithm. The usual convergence order of the regret function by a typical algorithm is $O(N\log{N})^{1/3}$ for MNB with bandit feedback in terms of the problem scale $N$. Our Thompson-Hedge algorithm feeds a set of sampled parameters to Hedge algorithm. In the proof of  Theorem \ref{theorem thompson-hedge algorithm}, it is shown that using the sampled parameters is ``as good as" using the true parameters under the Bayesian framework. Moreover, if the true parameters are known, then Hedge algorithm can solve the problem, which has a convergence order of $O(\log{N})$. Therefore, our Thompson-Hedge algorithm is supposed to retain a convergence rate of $O(\log{N})$ for the Bayesian sup regret in the special case.

Meanwhile, upper bound in Theorem \ref{theorem thompson-hedge algorithm} is also drawn in Figures \ref{fig11}-\ref{fig22}. It is worth noting that the upper bound in Theorem \ref{theorem thompson-hedge algorithm} holds uniformly on all possible values of $\theta$. Hence, it is concluded that when $\theta$ varies, the deviation of the regret from the upper bound also varies. In particular, as shown in the figures, the difference between the upper bound and the regret is comparatively large under $\theta=\theta_{1}$, $N=20$ while small or moderate under other three parameters. However, it can be observed from the figures that our upper bound is obviously sharper than that for the R.EXP3 algorithm in the sense that our upper bound is below the regret under R.EXP3 at some time points in all the four figures.

%%%%%%%%%%%%%%%%%%%%%%%%%%%%%%%%%%%%%%%%%%%%%%%%%%%%%%%%%%%%%%%%%%%%%%%%%%%%%%%%%%%%%%%%%%%%%%%%%%%%%%%
\subsection{Sensitivity to the variation}\label{subsection sensitivity to the variation}
\begin{figure}
\centering
\includegraphics[width=0.45\textwidth]{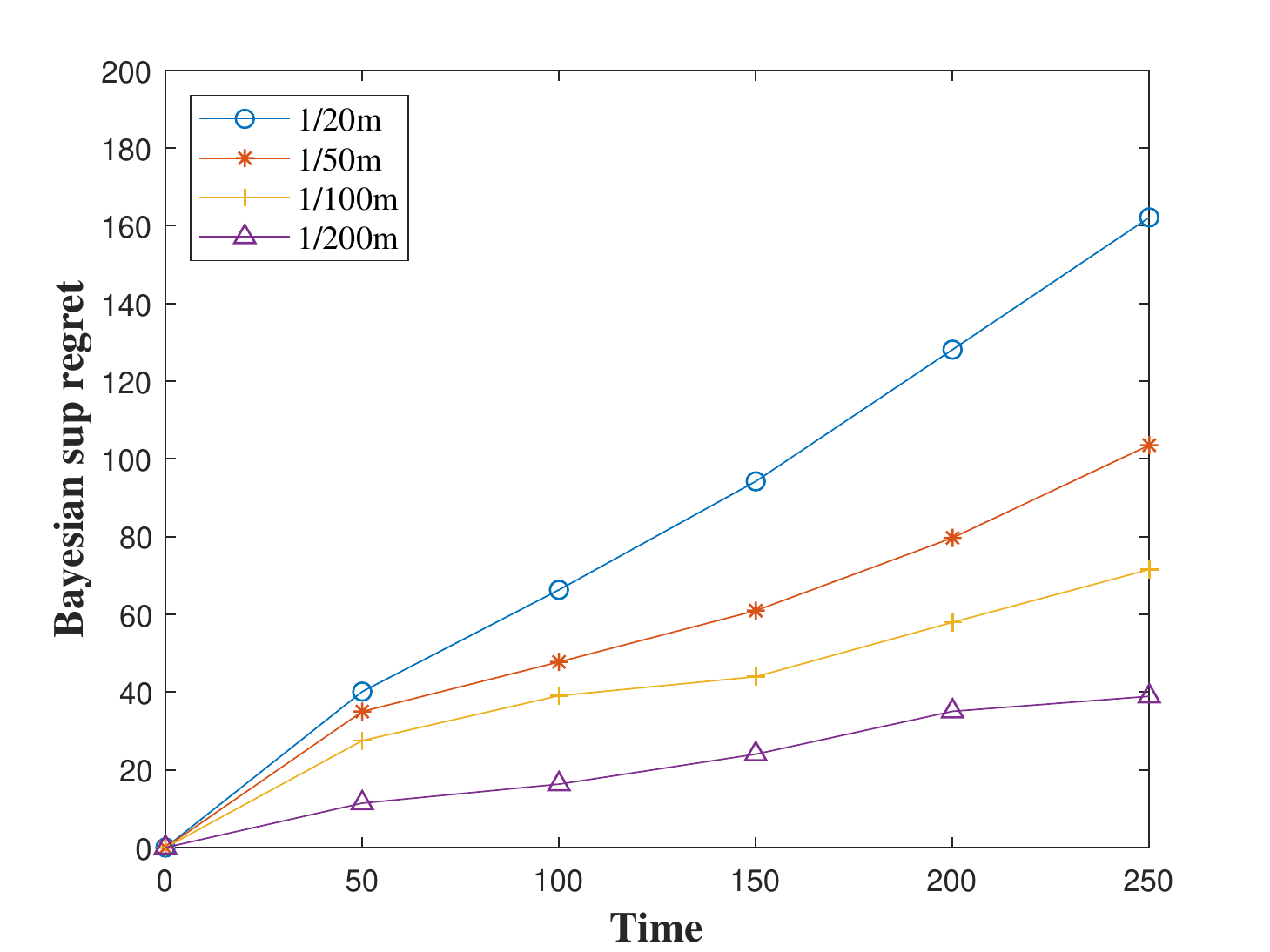}
\caption{Regrets under different variations ($\theta=\theta_{1}$, $N=20$).}
\label{fig31}
\end{figure}

\begin{figure}
\centering
\includegraphics[width=0.45\textwidth]{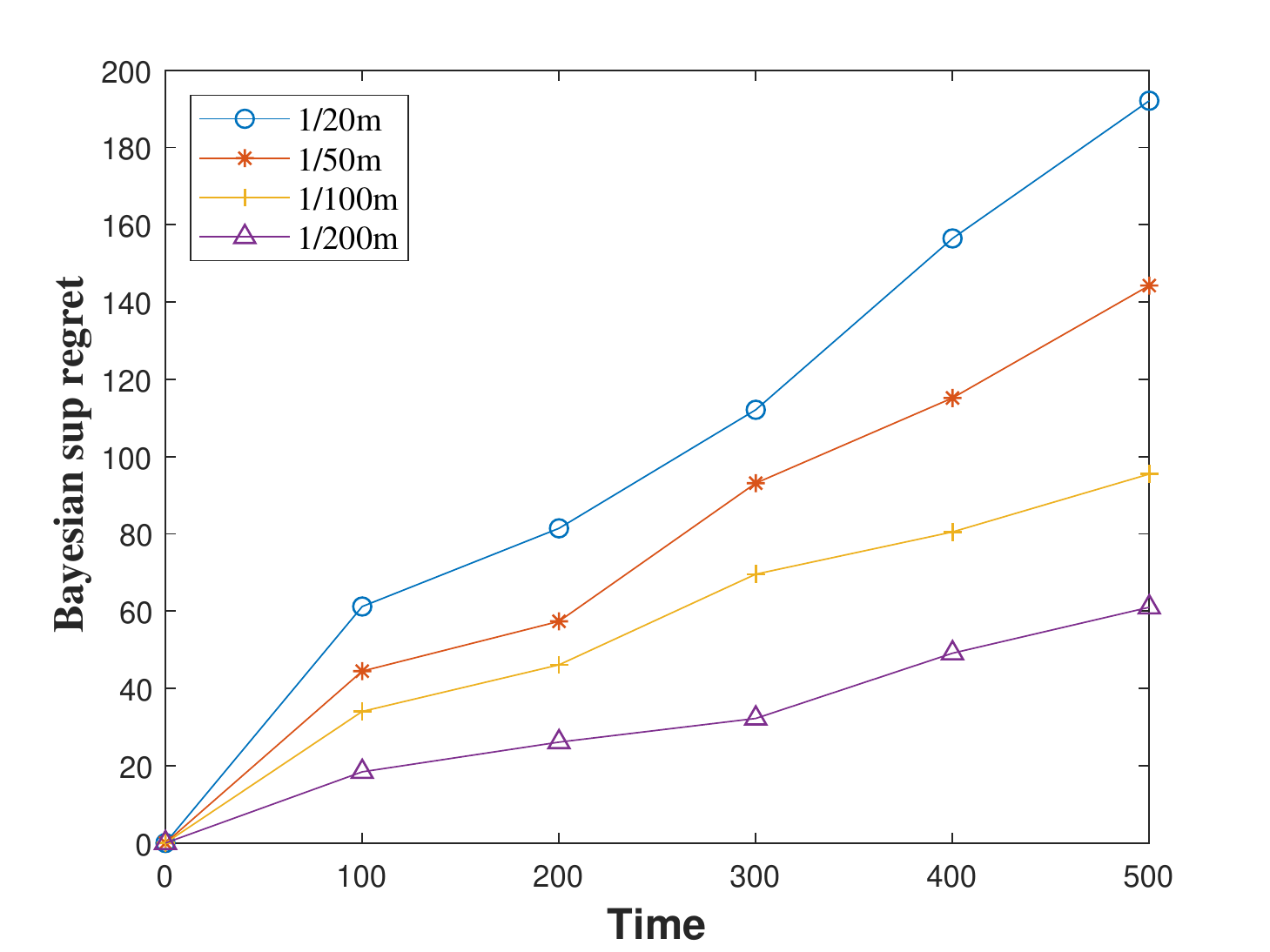}
\caption{Regrets under different variations ($\theta=\theta_{2}$, $N=20$).}
\label{fig32}
\end{figure}

This section presents the sensitivity analysis of our Thompson-Hedge algorithm to the variation constraint $\mathcal{V}_{T}$. In reality, according to the specific environment and workload under which the network functions, the adversary may generate the cost sequence subject to various rules. In our model, the variation constraint $\mathcal{V}_{T}$ is used to characterize the cost sequence. Thus, numerical results for the proposed Thompson-Hedge algorithm are presented under four levels of variation. In particular, the same steps given in Section \ref{subsection comparative study} are employed, but $4$ levels are selected for the variation scale in \textit{Step 2}: $1/20m$, $1/50m$, $1/100m$ and $1/200m$. Other parameters are set to be identical as those in Section \ref{subsection comparative study}. The simulation results are illustrated under both $\theta_{1}$ and $\theta_{2}$ as shown in Figures \ref{fig31} and \ref{fig32}.

%
%\begin{figure}[h]
%\begin{minipage}[t]{0.49\linewidth}
%\centering
%\includegraphics[width=\textwidth]{Figure_Sensitivity_1}
%\caption{Bayesian sup regret for $\theta=\theta_{1}$, $N=20$.}
%\label{fig31}
%\end{minipage}
%\begin{minipage}[t]{0.49\linewidth}
%\centering
%\includegraphics[width=\textwidth]{Figure_Sensitivity_2}
%\caption{Bayesian sup regret for $\theta=\theta_{2}$, $N=20$.}
%\label{fig32}
%\end{minipage}
%\end{figure}

The results in Figures \ref{fig31} and \ref{fig32} illustrate the different performances of our Thompson-Hedge algorithm under different variation scales. The Bayesian regret monotonically increases with the variation. This is because larger variation leads to more uncertainties of the cost sequence and makes the experience from historical observations unreliable. Thus, as an adaptive online learning algorithm, the Bayesian sup regret in our Thompson-Hedge algorithm may increase and the performance may have fluctuations. However, it should be noted that the incremental difference between the regret curves under $1/100m$ and $1/200m$ are not so significant compared to that between the curves under $1/50m$ and $1/20m$. This implies that when the variation is within $\mathcal{V}_{T}\leq T/50m$, the regret function is less sensitive to the variation than the situation that $\mathcal{V}_{T}$ is around $1/10m$. In the numerical example, the real data set (from Elia Grid, Belgium) is examined and it appears that the corresponding $\mathcal{V}_{T}$ is around $T/100m$. Therefore, it can be concluded that the sensitivity of our Thompson-Hedge algorithm to the variation may not be high in practice.

%%%%%%%%%%%%%%%%%%%%%%%%%%%%%%%%%%%%%%%%%%%%%%%%%%%%%%%%%%%%%%%%%%%%%%%%%%%%%%%%%%%%%%%%%%%%%%
\section{Concluding remarks}\label{section concluding remarks}
This study investigates the sequential control problem in modern energy and power systems with
smart grids. The existing work is extended by introducing an adversarial cost sequence with a variation
constraint. A Bayesian MNB model is constructed to cope with the problem and an online learning algorithm named Thompson-Hedge algorithm is proposed to retain a converging regret function. In addition, it is proved that the convergence rate of the regret in the proposed algorithm is superior to an existing algorithm that can be used for this problem.

In developing our algorithm, a basic idea is to firstly obtain sampled parameters from the posterior distribution when the reinforcement learning model is a partly parameterized Bayesian model. Subsequently, the sampled parameters are used instead of the true parameters in the model. It is worth pointing out that although specific models are used in this work, the paradigm of our algorithm framework can also be used for other classical models.

From a conceptual perspective, future research will explore the applications of the proposed algorithms to more situations in reality. For example, the proposed algorithm can be used to mitigate the impact of failure in tracking goods, or obtain suppliers to exchange inventory information due to successful cyber attacks. Other potential applications can be found in the domains of multi-robot systems \citep{liu2017distributed}, Internet of Things (IOT) \citep{perera2015emerging}, $e.g.$, traffic management, water distribution, waste management, online manufacturing and smart supply chain. It is also worth investigating the impact of other types of cyber attacks such as the botnet command and control, data exfiltration, data tampering, data destruction or even physical, destruction via alternation of critical software/data \citep{pillitteri2014guidelines} on the smart grid by properly modeling their influences on the Optimal Power Flow model according to their attack mechanisms. For example, the replay attacks maliciously repeat DR messages or dispatch commands, and the false data injection attacks purposely alter the integrity of a smart grid by compromising a subset of transmitted data packages and sending out inaccurate DR messages or dispatch commands.

%%%%%%%%%%%%%%%%%%%%%%%%%%%%%%%%%%%%%%%%%%%%%%%%%%%%%%%%%%%%%%%%%%%%%%%%%%%%%%%%%%%%%%%%%%%%%%%%%
%\newpage
\appendix
\renewcommand\thesection{\appendixname~\Alph{section}}
\renewcommand\theequation{\Alph{section}.\arabic{equation}}
\section{Proof of Lemma \ref{lemma hedge}}
In this section, detailed proof of Lemma \ref{lemma hedge} is provided, which is used in the proof of Theorem \ref{theorem thompson-hedge algorithm}. Before reaching the proof of Lemma \ref{lemma hedge}, a preliminary result is presented in the following lemma \cite{slivkins2019introduction}.

\begin{lemma}\label{lemma slivkins}[\cite{slivkins2019introduction}, Theorem 5.16]
If all per-time costs are in $[0,\,1]$ and $\varepsilon$ is chosen to be $\left(1-\sqrt{\ln{N}/2T}\right)^{-1}$, then the Hedge algorithm satisfies
\begin{align*}
&\max_{i\in\mathbb{N}}\left\{\sum_{t=\tau_{b-1}}^{\tau_{b}-1}K_{i,\,t}\cdot c_{i,\,t}
\right\}\\
&- \mathbf{E}^{H}\left(\sum_{t=\tau_{b-1}}^{\tau_{b}-1}K_{i_{t},\,t}\cdot c_{i_{t},\,t} \,\Big|\, \lambda,\,\vec{c}_{T}\right)\\
&\leq 2\sqrt{2}\cdot \sqrt{\Delta\ln{N}}.
\end{align*}
\end{lemma}
Note that the original theorem in \cite{slivkins2019introduction} concerns a reward-based problem and the parameter $\varepsilon$ is chosen to be $\sqrt{\ln{N}/2T}$. Actually, by using the transformation of
$cost=1-reward$ between the per-time cost and reward, the result can be transferred in the original theorem into the conclusion in Lemma \ref{lemma slivkins}. Detailed proof of Lemma \ref{lemma hedge} is given as follows.

%\begin{pf}
\textbf{\textup{Proof of Lemma \ref{lemma hedge}.}}
The proof is divided into four steps. In the first step, the best single action policy is used to decompose the regret function into two parts. In the second and third steps, upper bounds on the two parts from the first step are constructed, respectively. Finally, the upper bound on the original regret function is presented. Denote the sub-sequence of times when Hedge algorithm restarts as
$1=\tau_{0}\leq\tau_{1}\leq\cdots\leq\tau_{\left\lceil T/\Delta\right\rceil}$, which implies that
$\tau_{b+1}-\tau_{b}=\Delta$ for all $b=1,\cdots,\left\lceil T/\Delta\right\rceil-1$ and
$\tau_{\left\lceil T/\Delta\right\rceil}-\tau_{\left\lceil T/\Delta\right\rceil-1}\leq\Delta$. For all
$b=1,\cdots,\left\lceil T/\Delta\right\rceil$, note that the set of times $\left\{\tau_{b-1},\cdots,\tau_{b}-1\right\}$ exactly forms the $b$th batch.
 \\

\noindent\textbf{Step 1 Regret decomposition}.\\
For each $b=1,\cdots,\left\lceil T/\Delta\right\rceil$, the best single node $I_{b}^{*}$ in $b$th batch is defined as $I_{b}^{*}\triangleq {\arg\max}_{i\in\mathbb{N}}\left\{\sum_{t=\tau_{b-1}}^{\tau_{b}-1}\mu_{i}\cdot c_{i,\,t}\right\}$.
 A specific sequence $\{c_{i,\,t}\}_{t=1}^{T}\in\mathbb{C}_{T}$ is fixed and the regret function within $b$th batch is decomposed as
\begin{align}\label{A.1}
\begin{split}
& \sup_{\vec{c}_{T}\in\mathbb{C}_{T}}
\sum_{t=\tau_{b-1}}^{\tau_{b}-1}\mu_{i_{t}^{*}}\cdot c_{i_{t}^{*},\,t}-
\mathbf{E}^{H}\left(\sum_{t=\tau_{b-1}}^{\tau_{b}-1}\mu_{i_{t}}\cdot c_{i_{t},\,t} \,\Big|\, \lambda,\,\vec{c}_{T}\right)\\
&\leq
\sup_{\vec{c}_{T}\in\mathbb{C}_{T}}
\left\{\sum_{t=\tau_{b-1}}^{\tau_{b}-1}\mu_{i_{t}^{*}}\cdot c_{i_{t}^{*},\,t}-
\sum_{t=\tau_{b-1}}^{\tau_{b}-1}\mu_{I_{b}^{*}}\cdot c_{I_{b}^{*},\,t}\right\}\\
 &+\sup_{\vec{c}_{T}\in\mathbb{C}_{T}}
\Bigg\{\sum_{t=\tau_{b-1}}^{\tau_{b}-1}\mu_{I_{b}^{*}}\cdot c_{I_{b}^{*},\,t}\\
&-\mathbf{E}^{H}\left(\sum_{t=\tau_{b-1}}^{\tau_{b}-1}\mu_{i_{t}}\cdot c_{i_{t},\,t} \,\Big|\, \lambda,\,\vec{c}_{T}\right)\Bigg\}.
\end{split}
\end{align}
The two terms in the right hand side of (\ref{A.1}) are defined as
\begin{equation*}
R_{b,\,1}\triangleq
\sup_{\vec{c}_{T}\in\mathbb{C}_{T}}
\left\{\sum_{t=\tau_{b-1}}^{\tau_{b}-1}\mu_{i_{t}^{*}}\cdot c_{i_{t}^{*},\,t}-
\sum_{t=\tau_{b-1}}^{\tau_{b}-1}\mu_{I_{b}^{*}}\cdot c_{I_{b}^{*},\,t}\right\},
\end{equation*}
and
\begin{align*}
R_{b,\,2}\triangleq
&\sup_{\vec{c}_{T}\in\mathbb{C}_{T}}
\Bigg\{\sum_{t=\tau_{b-1}}^{\tau_{b}-1}\mu_{I_{b}^{*}}\cdot c_{I_{b}^{*},\,t}\\
&-\mathbf{E}^{H}\left(\sum_{t=\tau_{b-1}}^{\tau_{b}-1}\mu_{i_{t}}\cdot c_{i_{t},\,t} \,\Big|\, \lambda,\,\vec{c}_{T}\right)\Bigg\}.
\end{align*}
%
%%%%%%%%%%%%%%%%%%%%%%%%%%%%%%%%%%%%%%%%%%%%%%%%%%%%%%%%%%%%%%%%%%%%%%%%%%
\noindent\textbf{Step 2 Upper bounding $R_{b,\,1}$}. \\
Let
$\mathcal{V}_{T}^{b}\triangleq \sum_{t=\tau_{b-1}}^{\tau_{b}-2}\left|c_{i,\,t+1}-c_{i,\,t}\right|$ be the total variation of $c_{i,\,t}$ in $b$th batch. For all $\vec{c}_{T}\in\mathbb{C}_{T}$, the following relation holds
\begin{equation}\label{equation proof of lemma hedge 1}
\max_{\tau_{b-1}\leq t\leq \tau_{b}-1}
\left\{\mu_{i_{t}^{*}}\cdot c_{i_{t}^{*},\,t}-\mu_{I_{b}^{*}}\cdot c_{I_{b}^{*},\,t}\right\}
\leq 2m\mathcal{V}_{T}^{b}.
\end{equation}
By contradiction, if (\ref{equation proof of lemma hedge 1}) does not hold, then there is at least one time
$\tau_{b-1}\leq t_{0}\leq \tau_{b}-1$ such that
$\mu_{i_{t}^{*}}\cdot c_{i_{t}^{*},\,t_{0}}-\mu_{I_{b}^{*}}\cdot c_{I_{b}^{*},\,t_{0}}\geq 2m\mathcal{V}_{T}^{b}$.
Let $i_{t_{0}}^{*}=i_{0}$. Since $\mu_{i}\leq m$, $\forall i\in\mathbb{N}$, for all $\tau_{b-1}\leq t\leq \tau_{b}-1$, it follows
\begin{align*}
\mu_{i_{0}}\cdot c_{i_{0},\,t}
&> \mu_{i_{0}}\left(c_{i_{0},\,t_{0}} - \mathcal{V}_{T}^{b}\right)
> \mu_{i_{0}}\cdot c_{i_{0},\,t_{0}} - m\mathcal{V}_{T}^{b}\\
&> \mu_{I_{b}^{*}}\cdot c_{I_{b}^{*},\,t_{0}} + m\mathcal{V}_{T}^{b}
> \mu_{I_{b}^{*}}\left(c_{I_{b}^{*},\,t_{0}} + \mathcal{V}_{T}^{b}\right)\nonumber\\
&> \mu_{I_{b}^{*}}\cdot c_{I_{b}^{*},\,t}.
\end{align*}
However, this contradicts with the fact that $I_{b}^{*}$ is the optimal single node of $b$th batch. Thus, inequality (\ref{equation proof of lemma hedge 1}) holds. Therefore, it can be obtained that
\begin{align*}
R_{b,\,1}
& =
\sup_{\vec{c}_{T}\in\mathbb{C}_{T}}\sum_{t=\tau_{b-1}}^{\tau_{b}-1}
\left(\mu_{i_{t}^{*}}\cdot c_{i_{t}^{*},\,t}-\mu_{I_{b}^{*}}\cdot c_{I_{b}^{*},\,t}\right)\nonumber\\
&\leq
\Delta\sup_{\vec{c}_{T}\in\mathbb{C}_{T}}\max_{\tau_{b-1}\leq t\leq \tau_{b}-1}
\left\{\mu_{i_{t}^{*}}\cdot c_{i_{t}^{*},\,t}-\mu_{I_{b}^{*}}\cdot c_{I_{b}^{*},\,t}\right\}\nonumber\\
& \leq
2\Delta m \mathcal{V}_{T}^{b}.
\end{align*}
%
%%%%%%%%%%%%%%%%%%%%%%%%%%%%%%%%%%%%%%%%%%%%%%%%%%%%%%%%%%%%%%%%%%%%%%%%%%%%%%%%%%%%
\noindent\textbf{Step 3 Upper bounding $R_{b,\,2}$}.\\
Note that
\begin{align*}
\sum_{t=\tau_{b-1}}^{\tau_{b}-1}\mu_{I_{b}^{*}}\cdot c_{I_{b}^{*},\,t}
&=\sum_{t=\tau_{b-1}}^{\tau_{b}-1}\mathbf{E}\left(K_{I_{b}^{*},\,t}\cdot c_{I_{b}^{*},\,t}\right)\nonumber\\
&=\max_{i\in\mathbb{N}}\left\{\sum_{t=\tau_{b-1}}^{\tau_{b}-1}\mathbf{E}\left(K_{i,\,t}\cdot c_{i,\,t}\right)\right\}\nonumber\\
&\leq
\mathbf{E}\left(\max_{i\in\mathbb{N}}\left\{\sum_{t=\tau_{b-1}}^{\tau_{b}-1}K_{i,\,t}\cdot c_{i,\,t}
\right\}\right).
\end{align*}

Therefore, it holds that
\begin{align*}
R_{b,\,2}
& \leq\sup_{\vec{c}_{T}\in\mathbb{C}_{T}}\Bigg\{
\mathbf{E}\Bigg[\left(\max_{i\in\mathbb{N}}\left\{\sum_{t=\tau_{b-1}}^{\tau_{b}-1}K_{i,\,t}\cdot c_{i,\,t}
\right\}\right)\\
&- \mathbf{E}^{H}\left(\sum_{t=\tau_{b-1}}^{\tau_{b}-1}\mu_{i_{t}}\cdot c_{i_{t},\,t} \,\Big|\, \lambda,\,\vec{c}_{T}\right)\Bigg]\Bigg\}\nonumber\\
& \leq\sup_{\vec{c}_{T}\in\mathbb{C}_{T}}\Bigg\{
\mathbf{E}\Bigg[\left(\max_{i\in\mathbb{N}}\left\{\sum_{t=\tau_{b-1}}^{\tau_{b}-1}K_{i,\,t}\cdot c_{i,\,t}
\right\}\right)\\
&- \mathbf{E}^{H}\left(\sum_{t=\tau_{b-1}}^{\tau_{b}-1}k_{i_{t},\,t}\cdot c_{i_{t},\,t} \,\Big|\, \lambda,\,\vec{c}_{T}\right)\Bigg]\Bigg\}.
\end{align*}
According to Lemma \ref{lemma hedge} and noting that the relation in Lemma \ref{lemma hedge} holds for arbitrary $\vec{c}_{T}$ satisfying $K_{i,\,t}\cdot c_{i,\,t}\leq 1$, $\forall i,\,t$, it follows
\begin{align*}
R_{b,\,2}\leq 2\sqrt{2}\cdot \sqrt{\Delta\ln{N}}.
\end{align*}
%
%%%%%%%%%%%%%%%%%%%%%%%%%%%%%%%%%%%%%%%%%%%%%%%%%%%%%%%%%%%%%%%%%%%%%%%%%%
\noindent\textbf{Step 4 Upper bounding the sup regret}.\\
In the final step, discussions in the previous two steps and the upper bound on the sup regret by Hedge algorithm are summarized. Note that
\begin{align}\label{A.3}
&\sup_{\vec{c}_{T}\in\mathbb{C}_{T}} R^{H}\left(\lambda,\,\vec{c}_{T},\,T\right)
 \leq
\sum_{b=1}^{\left\lceil T/\Delta\right\rceil}\left(R_{b,\,1}+R_{b,\,2}\right)\nonumber\\
&\leq \sum_{b=1}^{\left\lceil T/\Delta\right\rceil}
\left(2\Delta m\mathcal{V}_{T}^{b} + 2\sqrt{2}\cdot \sqrt{\Delta\ln{N}}\right)\nonumber\\
& \leq
2\Delta m\mathcal{V}_{T} + \left(\frac{T}{\Delta}+1\right)2\sqrt{2}\cdot \sqrt{\Delta\ln{N}},
\end{align}
where the third inequality follows from the relation
$\sum_{b=1}^{\left\lceil T/\Delta\right\rceil}\mathcal{V}_{T}^{b}=\mathcal{V}_{T}$.

Embedding $\Delta=\left\lceil\left(\ln{N}\right)^{1/3}\left({T}/{m\mathcal{V}_{T}}\right)^{2/3}\right\rceil$ into (\ref{A.3}), it follows
\begin{align*}
&\sup_{\vec{c}_{T}\in\mathbb{C}_{T}} R^{\pi}\left(\lambda,\,\vec{c}_{T},\,T\right)\\
&\leq
2\left[\left(\ln{N}\right)^{1/3}\left(\frac{T}{m\mathcal{V}_{T}}\right)^{2/3}+1\right] m\mathcal{V}_{T}\\
&+ 2\sqrt{2}\cdot T\cdot\sqrt{\ln{N}\left(\ln{N}\right)^{-1/3}\left(\frac{T}{m\mathcal{V}_{T}}\right)^{-2/3}}
\nonumber\\
& + 2\sqrt{2}\sqrt{\ln{N}\left[\left(\ln{N}\right)^{1/3}\left(\frac{T}{m\mathcal{V}_{T}}\right)^{2/3}+1\right]}.
\end{align*}
Since $N>2$, Assumption \ref{assumption cost} leads to $\left(\ln{N}\right)^{1/3}\left({T}/{m\mathcal{V}_{T}}\right)^{2/3}>1$ and
$\left[\ln{N}/T(m\mathcal{V}_{T})^{2}\right]<1$. It follows
\begin{align*}
&\sup_{\vec{c}_{T}\in\mathbb{C}_{T}} R^{\pi}\left(\lambda,\,\vec{c}_{T},\,T\right)\nonumber\\
& \leq
\left(4+2\sqrt{2}\right)\left(m\mathcal{V}_{T}\ln{N}\right)^{1/3}\left(T\right)^{2/3}\nonumber\\
&+ 4\left(\ln{N}\right)^{2/3}\left(\frac{T}{m\mathcal{V}_{T}}\right)^{1/3}\nonumber\\
& \leq
\left(4+2\sqrt{2}\right)\left(m\mathcal{V}_{T}\ln{N}\right)^{1/3}\left(T\right)^{2/3}\nonumber\\
&+ 4\left(m\mathcal{V}_{T}\ln{N}\right)^{1/3}\left(T\right)^{2/3}
\left[\frac{\ln{N}}{T\left(m\mathcal{V}_{T}\right)^{2}}\right]^{1/3}\nonumber\\
& \leq
\left(4+2\sqrt{2}\right)\left(m\mathcal{V}_{T}\ln{N}\right)^{1/3}\left(T\right)^{2/3}\nonumber\\
&+ 4\left(m\mathcal{V}_{T}\ln{N}\right)^{1/3}\left(T\right)^{2/3}\nonumber\\
& \leq
\left(8+2\sqrt{2}\right)\left(m\mathcal{V}_{T}\ln{N}\right)^{1/3}\left(T\right)^{2/3},
\end{align*}
which concludes the proof.
%\end{pf}

\bibliographystyle{agsm}        % Include this if you use bibtex
\bibliography{reference}           % and a bib file to produce the
                                 % bibliography (preferred). The
                                 % correct style is generated by
                                 % Elsevier at the time of printing.

%\begin{thebibliography}{99}     % Otherwise use the
                                 % thebibliography environment.
                                 % Insert the full references here.
                                 % See a recent issue of Automatica
                                 % for the style.
%  \bibitem[Heritage, 1992]{Heritage:92}
%     (1992) {\it The American Heritage.
%     Dictionary of the American Language.}
%     Houghton Mifflin Company.
%  \bibitem[Able, 1956]{Abl:56}
%     B.~C.~Able (1956). Nucleic acid content of macroscope.
%     {\it Nature 2}, 7--9.
%  \bibitem[Able {\em et al.}, 1954]{AbTaRu:54}
%     B.~C. Able, R.~A. Tagg, and M.~Rush (1954).
%     Enzyme-catalyzed cellular transanimations.
%     In A.~F.~Round, editor,
%     {\it Advances in Enzymology Vol. 2} (125--247).
%     New York, Academic Press.
%  \bibitem[R.~Keohane, 1958]{Keo:58}
%     R.~Keohane (1958).
%     {\it Power and Interdependence:
%     World Politics in Transition.}
%     Boston, Little, Brown \& Co.
%  \bibitem[Powers, 1985]{Pow:85}
%     T.~Powers (1985).
%     Is there a way out?
%     {\it Harpers, June 1985}, 35--47.

%\end{thebibliography}

%\appendix
%\section{A summary of Latin grammar}    % Each appendix must have a short title.
%\section{Some Latin vocabulary}         % Sections and subsections are supported
%                                        % in the appendices.
\end{document}